\documentclass[final,5p,times,twocolumn]{elsarticle}
\usepackage{amsmath}
\usepackage{xcolor}
\usepackage{accents}
\graphicspath{{figs/}}

\usepackage[switch]{lineno}
\usepackage[math]{cellspace}
\usepackage{mathtools}
\usepackage{inst_paper}
\usepackage{hyperref}
\journal{Classical and Quantum Gravity}
\usepackage{autofigs}

\begin{document}
\begin{frontmatter}

\title{The Holometer: An Instrument to Probe Planckian Quantum Geometry}

\author[fnal]{Aaron Chou}
\ead{achou@fnal.gov}

\author[fnal]{Henry Glass}
\ead{glass@fnal.gov}

\author[um]{H. Richard Gustafson}
\ead{gustafso@umich.edu}

\author[fnal,uc]{Craig Hogan}
\ead{cjhogan@fnal.gov}

\author[fnal,uc,vand]{Brittany L. Kamai}
\ead{bkamai@ligo.caltech.edu}

\author[kaist]{Ohkyung Kwon}
\ead{o.kwon@kaist.ac.kr}

\author[mit]{Robert Lanza}
\ead{bobbylanza@gmail.com}

\author[mit]{Lee McCuller}
\ead{lee.mcculler@gmail.com}

\author[uc]{Stephan S. Meyer\corref{cor1}}
\ead{meyer@uchicago.edu}

\author[uc,um]{Jonathan Richardson}
\ead{jonathan.richardson@uchicago.edu}

\author[fnal]{Chris Stoughton}
\ead{stoughto@fnal.gov}

\author[fnal]{Ray Tomlin}
\ead{tomlin@fnal.gov}

\author[mit]{Rainer Weiss}
\ead{weiss@ligo.mit.edu}

\cortext[cor1]{Corresponding Author}
\address[fnal]{Fermi National Accelerator Laboratory}
\address[um]{University of Michigan}
\address[uc]{University of Chicago}
\address[cit]{California Institute of Technology}
\address[kaist]{Korea Advanced Institute of Science and Technology}
\address[mit]{Massachusetts Institute of Technology}
\address[vand]{Vanderbilt University}

\begin{abstract}
This paper describes the Fermilab Holometer, an instrument for measuring correlations of position variations over a four-dimensional volume of space-time. The apparatus consists of two co-located, but independent and isolated, 40-m power-recycled Michelson interferometers, whose outputs are cross-correlated to 25~MHz. The data are sensitive to correlations of differential position across the apparatus over a broad band of frequencies up to and exceeding the inverse light crossing time, 7.6~MHz. A noise model constrained by diagnostic and environmental data distinguishes among physical origins of measured correlations, and is used to verify shot-noise-limited performance. These features allow searches for exotic quantum correlations that depart from classical trajectories at spacelike separations, with a strain noise power spectral density sensitivity smaller than the Planck time. The Holometer in current and future configurations is projected to provide precision tests of a wide class of models of quantum geometry at the Planck scale, beyond those already constrained by currently operating gravitational wave observatories.
\end{abstract}

\end{frontmatter}

\twocolumn
\tableofcontents

\newcommand{\abs}[1]{\left|#1\right|}
\newcommand{\abssq}[1]{\abs{#1}^2}
\newcommand{\holoMag}{\Xi}
\newcommand{\Wholo}{\sop{W}\tu{exotic}}

\newcommand{\ifoL}[1][]{\tb{1#1}}
\newcommand{\ifoT}[1][]{\tb{2#1}}

\newcommand{\conv}{\mathop{*}}
\newcommand{\vconv}{\mathop{\mathbf{*}}}
\newcommand{\vect}[1]{\mathbf{#1}}
\newcommand{\matr}[1]{\mathcal{\mathbf{#1}}}
\newcommand{\conj}[1]{\overline{#1}}

\newcommand{\tu}[1]{^\text{#1}}
\newcommand{\tb}{_\text}
\newcommand{\sop}[1]{\hat{#1}}
\newcommand{\vsop}[1]{\sop{\vect{#1}}}

\newcommand{\WBG}{ \vsop{W}\tu{env}}
\newcommand{\WBGL}{\vsop{W}\tu{sys}\ifoL}
\newcommand{\WBGT}{\vsop{W}\tu{sys}\ifoT}

\newcommand{\VBGL}[1][]{{\vect{\Lambda}\ifoL\tu{#1}}}
\newcommand{\VBGT}[1][]{{\vect{\Lambda}\ifoT\tu{#1}}}
\newcommand{\VNL}[1][]{ {\vect{ \Gamma}\ifoL\tu{#1}}}
\newcommand{\VNT}[1][]{ {\vect{ \Gamma}\ifoT\tu{#1}}}
\newcommand{\Vdot}{\cdot}
\newcommand{\CSD}[1]{\text{CSD}\left[#1\right]}
\newcommand{\PSD}[1]{\text{PSD}\left[#1\right]}
\newcommand{\Arg}[1]{\text{ARG}\left[#1\right]}
\newcommand{\dft}[1]{\mathcal{F}\left[#1\right]}

\newcommand{\W}{\sop{W}}
\newcommand{\Wv}[1][]{\vsop{W}\tu{#1}}

\newcommand{\stau}{\sop{\tau}}

\newcommand{\michdl}[1][]{^{\delta\hspace{-.07em}L\text{#1}}}
\newcommand{\michas}[1][]{^{\text{ASPD}\text{#1}}}
\newcommand{\CSDjr}[3][]{\text{CSD}\left[#2,#3 #1\right]}
\newcommand{\PSDjr}[1]{\text{PSD}\left[#1\right]}
\newcommand{\Src}[1][]{{S}_{#1}}
\newcommand{\Dinj}[0]{\sop{D}}
\newcommand{\Minj}[0]{\sop{M}^\mathrm{inj}}
\newcommand{\Vinj}[0]{\sop{V}^\mathrm{inj}}
\newcommand{\HdaqHF}[1][]{H^\mathrm{DAQ}_{#1{\rm HF}}}
\newcommand{\HdaqLF}[1][]{H^\mathrm{DAQ}_{#1{\rm LF}}}
\newcommand{\hHdaqHF}[1][]{\sop{H}^\mathrm{DAQ}_{#1{\rm HF}}}
\newcommand{\hHdaqLF}[1][]{\sop{H}^\mathrm{DAQ}_{#1{\rm LF}}}
\newcommand{\Hifo}[1][]{H^\mathrm{IFO}_{#1}}
\newcommand{\HcalHF}[1][]{H\michdl_{#1{\rm HF}}}
\newcommand{\hHcalHF}[1][]{\sop{H}\michdl_{#1 {\rm HF}}}
\newcommand{\HcalLF}[1][]{H\michdl_{#1{\rm LF}}}
\newcommand{\hHcalLF}[1][]{\sop{H}\michdl_{#1{\rm LF}}}
\newcommand{\VHF}[1][]{{V}\michas_{#1{\rm HF}}}
\newcommand{\hVHF}[1][]{\sop{{V}}\michas_{#1{\rm HF}}}
\newcommand{\VLF}[1][]{{V}\michas_{#1{\rm LF}}}
\newcommand{\hVLF}[1][]{\sop{{V}}\michas_{#1{\rm LF}}}
\newcommand{\Vac}[1][]{{V}^\mathrm{AC}_{#1}}
\newcommand{\hVac}[1][]{\sop{{V}}^\mathrm{AC}_{#1}}
\newcommand{\Vdc}[1][]{{V}^\mathrm{DC}_{#1}}
\newcommand{\hVdc}[1][]{\sop{{V}}^\mathrm{DC}_{#1}}
\newcommand{\Vref}[0]{{V}^\mathrm{ref}}
\newcommand{\hVref}[0]{\sop{{V}}^\mathrm{ref}}
\newcommand{\Hled}[0]{H^\text{LED}}
\newcommand{\Galign}[1][]{G^\mathrm{#1align}}
\newcommand{\HadcHF}[1][]{H^\mathrm{ADC}_{#1HF}}
\newcommand{\HadcLF}[1][]{H^\mathrm{ADC}_{#1LF}}
\newcommand{\HpdHF}[1][]{H^\mathrm{PD}_{#1HF}}
\newcommand{\HpdLF}[1][]{H^\mathrm{PD}_{#1LF}}
\newcommand{\Href}[0]{H^\mathrm{ref}}
\newcommand{\hHadcHF}[1][]{\sop{H}^\mathrm{ADC}_{#1HF}}
\newcommand{\hHadcLF}[1][]{\sop{H}^\mathrm{ADC}_{#1LF}}
\newcommand{\hHpdHF}[1][]{\sop{H}^\mathrm{PD}_{#1HF}}
\newcommand{\hHpdLF}[1][]{\sop{H}^\mathrm{PD}_{#1LF}}
\newcommand{\hHref}[0]{\sop{H}^\mathrm{ref}}
\newcommand{\Pshot}[0]{{P}^\mathrm{shot}}
\newcommand{\hPshot}[0]{\sop{M}^\mathrm{shot}}
\newcommand{\Zac}[0]{Z^\mathrm{AC}}
\newcommand{\estZac}[0]{Z^\mathrm{AC}}
\newcommand{\Zdc}[0]{Z^\mathrm{DC}}
\newcommand{\estZdc}[0]{Z\tu{DC}}

\newcommand*{\bdot}[1]{\accentset{\mbox{\large\bfseries .}}{#1}}

\section{Introduction}
    \label{sec:intro}

A possible consequence of quantum gravity is the creation and annihilation of high-energy virtual particles that gravitationally alter space-time to be ``foamy'' or ``fuzzy'' at the Planck scale. Various models mathematically suggest such nonclassicality \cite{Wheeler:1957, Hawking:1978, Hawking:1980, Ashtekar:1992, Ellis:1992}, but it is not clear exactly how much the physical metric departs from the classical structure, nor how the effects of Planck-scale fluctuations affect the geodesic of an object on larger scales. Some estimates of the quantum deviations from a classical background \cite{Hogan:2008a, Hogan:2008b, Hogan:2009, Hogan:2012, Hogan:2013, Kwon:2014, Hogan:2015a, Hogan:2015b, Hogan:2016}, based on extrapolations of standard physics normalized to match black hole entropy, have suggested that quantum fluctuations could lead to positional decoherence on small but measurable scales ($10^{-18}$ m in a 1-m system). Moreover, these fluctuations of relative position are predicted to correlate spatially over distances small compared to the system size.

Laser interferometry is an experimental technique well-suited for testing theories of positional decoherence. In recent years, large interferometers, developed primarily to detect gravitational waves \cite{Adhikari:2014}, have crossed a remarkable threshold in precision measurement of the relative space-time positions of massive bodies (specifically, mirrors) at macroscopic separation. The sensitivity of an interferometer can be expressed as a strain noise power spectral density,
\begin{align}
\label{eq:powersd}
h^2(f, t) &\equiv \int_{-\infty}^{\infty} \left\langle \frac{\delta\hspace{-.1em}L(t)}{L} \, \frac{\delta\hspace{-.1em}L(t - \tau)}{L} \right\rangle e^{- 2\pi i \tau f} \, d\tau \;,
\end{align}
the spectral variance of apparent fractional position fluctuations $\delta L/L$ for mirrors separated from a center optic by distance $L$\footnote{In Eq.~\ref{eq:powersd}, brackets indicate the expectation over the distribution of the random signal $\delta\hspace{-.1em}L(t)$, which need not be stationary. For a stationary signal, Eq.~\ref{eq:powersd} reduces to an explicitly time-independent form, $h^2(f,t)=h^2(f)$.}. Interferometric measurements of this noise density, which has units of time, can now achieve sensitivity below the Planck time, $t_P\equiv \sqrt{\hbar G/c^5} = 5.391\times 10^{-44}$ s. Thus there is an opportunity for experiments to search for possible quantum departures from classical space-time, by measuring with Planck precision the coherence of world lines on large scales.

Existing gravitational wave detectors, such as LIGO, Virgo, and GEO600, constrain some hypotheses about quantum-geometrical decoherence, particularly those based on fluctuations of the metric \citep{Kwon:2014}. However, since those detectors are specifically optimized to measure a metric strain at frequencies small compared their inverse light travel time, a much larger space of possible effects originating from Planck-scale fluctuations remains unexplored. The Fermilab Holometer is the first instrument built for experimental studies of exotic correlations of macroscopic trajectories through space-time. It consists of two co-located, but independent and isolated, 40-m power-recycled Michelson interferometers, whose outputs are sampled faster than the light crossing time. Correlations, determined by the layout of the two interferometers, are measured in the time dimension represented by the signal streams at both spacelike and timelike separations. As a dedicated experiment, the Holometer is projected to provide Planck-precision tests of a wider class of quantum decoherence theories, significantly extending constraints from currently operating gravitational wave observatories.

This paper describes the instrumentation of the dual-interferometer Holometer apparatus. Although it is based largely on technology developed for gravitational wave detectors, the Holometer differs in several important respects. The high-level principles governing its design are:
\begin{enumerate}
\item 
It must be able to measure or constrain universal exotic correlations with $h^2$ much smaller than $t_P$, and distinguish them from conventional environmental sources.
\item
It must be able to measure cross correlations between systems that are isolated from each other in all physical respects, except for their spatial and temporal relationships. Thus, it must be possible for interferometers to be close together, but to  have separate and isolated input and output optics, vacuum systems, electronics, clocks, and data streams.
\item
The light paths must compare multiple directions in space, and insofar as practical, be reconfigurable to explore different possible spatial patterns of correlation.
\item
The time resolution of  time domain correlation measurements must be much smaller than a light crossing time. As a corollary, the bandwidth of the frequency domain spectrum must be significantly larger than an inverse light crossing time (or free spectral range). This requirement demands a high bandwidth and data sampling rate.
\item
The frequency resolution of  power spectral density measurements must be sufficient to isolate environmental sources and provide system diagnostics.
\item
It must have a simple optical design, with propagation in vacuo as much as possible, to allow straightforward interpretation in terms of universal properties of space.
\item
Cost constrains the design, particularly the size of the system.
\end{enumerate}
The following sections elaborate on how these principles guided various design choices. They lead to some major design differences from gravitational wave observatories: a smaller apparatus, higher sampling frequency and bandwidth, and simpler mechanical systems. In many other respects, such as optics, cavity design, and control systems, the design is remarkably similar.

Initial data for the first 145~hours of integration of the first-generation experiment are presented in \S\ref{sec:data-integration}. After environmental background analysis, described in \S\ref{sec:env_background}, the non-vetoed data are consistent with zero correlation. The presentation herein does not test for any particular quantum-geometrical noise model, as it does not weight the data by any model noise spectrum. Scientific analysis of this data set is separately performed in \cite{HoloPRL} to place constraints on a candidate model. Final analysis of the full 704-hour data set of the first-generation Holometer, which has the statistical power to constrain a broader class of models, is forthcoming.

\section{Interferometer Design \& Optical System}

Each of the two interferometers converts position fluctuations into a changing optical power incident on a photodetector at the anti-symmetric (AS) port, or ``dark'' port, which is typically operated at destructive interference. The optical power at the AS port is 
\begin{align}
P_\text{AS} &= P_{\rm BS}
  \left(\epsilon_\text{cd} + (1-2\epsilon_\text{cd})
    \sin^2\bigg(\frac{2\pi}{\lambda} \delta\hspace{-.1em}L\bigg)
  \right)
  \label{eq:michelson_AS}
\end{align}
for an optical power $P_{\rm BS}$ incident on the beamsplitter. The term $\delta\hspace{-.1em}L = L_x - L_y$ represents the differential arm length, although it will be used herein to refer to simultaneous local displacements of the end mirrors. This distinction is relevant for defining signals sampled up to the light-crossing time. The term $\epsilon_\text{cd}$ is the fraction of so-called contrast-defect light, where the gain mismatch or transverse mode shape fails to perfectly interfere at the beamsplitter (``BS'') and is not modulated by $\delta\hspace{-.1em}L$. It is characterized in \S\ref{sec:defect_performance}.

The measurement of $\delta\hspace{-.1em}L$ via $P\tb{AS}$ is limited by Poisson statistics, or ``shot noise,'' in estimating the mean incidence rate of photons as an optical power. For an average photon rate of $\bdot{\mathcal{N}} = \frac{d\mathcal{N}}{dt}$, the spectral density of shot noise scales as $\bdot{\mathcal{N}}^{-1/2}$. A quantum-geometrical effect would appear as correlated deviations from the mean incidence rate on time scales shorter than the round-trip light-crossing time of the apparatus. Using notation elaborated in \S \ref{sec:Characterization}, the shot noise-limited sensitivity to differential end-mirror displacements can be expressed as
\begin{align}
  \PSD{\sop{M}\michdl; f, t} &\ge {2 \left(\frac{\lambda}{4\pi}\right)}^2 \frac{1}{\bdot{\mathcal{N}}(t)} & \text{using } \bdot{\mathcal{N}}(t) &= \frac{P_\text{BS}(t)}{h \nu} 
  \label{eq:shot_noise_limited_sensitivity}
\end{align}
where $\PSD{\cdot}$ provides the expectation of the \emph{one-sided} power-spectrum. Increasing the optical power can thus significantly improve the sensitivity of the measurement, which the Holometer achieves through the addition of a power recycling mirror at each interferometer's input port. This mirror forms a Fabry-Perot cavity with the end mirrors, resulting in a resonant enhancement of the optical field strength inside the instrument. The static field behavior of the interferometer can be modeled as the linear system depicted in Fig.~\ref{fig:model_signal_flow}.
\autofiguresvg{
  folder=figs/model/,
  file=model_signal_flow, 
  caption = {
    Signal flow diagram of the interferometer static fields, representing the linear system of equations for the interferometer response. The fields are sourced by $E\tb{laser}$. Each reflection and transmission coefficient is given by $r$ or $t$, subscripted by its respective optic. $F_\lambda$ is the optical carrier frequency and the arm lengths are $L_x$ and $L_y$. The round nodes represent internal states of the physical system whereas the square nodes are output fields observed in photodetectors, such as the anti-symmetric port field $E\tb{AS}$. To fully model the contrast-defect, additional copies of this diagram must be added to represent additional transverse modes, with small transfer coefficients arising by defects in the end mirror and beamsplitter shapes. The signal sidebands from the end-mirrors are signified with red and blue arrows, which source (unshown) copies of the graph at the modulation frequencies. At each port the optical field at carrier and sideband frequencies beat together to model the full frequency-dependent response.
  },
}

\subsection{Power-Recycling Cavity}

The Michelson interferometer, formed by the beamsplitter and two end mirrors, forms an effective mirror where the losses are determined by the fraction of light escaping to the AS port and the remainder is reflected. The addition of a power recycling mirror (PRM) forms a cavity with this effective mirror. The power-recycled interferometers are designed to be nearly confocal resonators, folded by the $45^\circ$ incidence beamsplitter so that each arm forms a flat-curved cavity. They have an arm length of $L= 39.2$~m and an end mirror radius of curvature of $R=75.1$~m, which is matched between the two arms to within 10~cm. The resulting waist, with radius
\begin{align}
w_0 = \sqrt{\frac{\lambda}{2\pi}\sqrt{2L(2R - 2L)}} \approx 3.57 \ \mathrm{mm}
\end{align}
lays at the position of the flat PRM. The end mirrors are each located nearly one Rayleigh range away, 
\begin{align}
Z_R &\equiv \frac{\pi w_0^2}{\lambda} = 37.6 \ \mathrm{m} 
\end{align}
where the beam half-width has grown to $w_1\approx5$~mm. The deviation from a pure confocal configuration satisfies the resonator stability criterion, $R< 2L$ (for a review of laser resonators, see \citep{kogelnik_li}).

This configuration additionally avoids co-resonances of higher-order cavity modes (HOM) with the fundamental mode. The Gouy phase shift for one-way propagation down the interferometer arm is
\begin{align}
\phi_\mathrm{Gouy} &\equiv \arctan{\left(\frac{L}{Z_R}\right)} = 0.806\;{\rm rad} \;(46.2^\circ)
\label{eq:gouy-shift}
\end{align}
Any Hermite-Gauss mode, $H_{mn}$, attains a round-trip phase excess of
\begin{align}
\phi_\mathrm{mn} = 2(1+m+n) \, \phi_\mathrm{Gouy}
\label{eq:hom_freqs}
\end{align}
compared to the propagation of unfocused spherical wavefronts. When $2(m+n)\phi_\mathrm{Gouy}$ is an integer multiple of
$2\pi$, the $H_{mn}$ mode will be co-resonant with the fundamental $H_{00}$ mode. With $1.61$~rad ($92^\circ$) of phase separation between each mode order, the lower-order HOM resonances are well-separated from the fundamental. The fourth-order mode wraps back near the fundamental, but the $0.17$~rad ($10^\circ$) phase separation is ${>}100$~times the cavity linewidth.

The power recycling cavity provides not only the resonant power enhancement, but also filters noise sidebands from the laser. Because of the higher-order-mode separation, different transverse modes have different filtered noise spectra, which beat with residual light at the Michelson anti-symmetric port readout. The modes then show as noise peaks at the HOM resonance frequencies in the readout spectra, confirming estimates of the arm length and Gouy separation. These noise measurements are described in \S\ref{sec:am-mitigation}.

\subsection{Instrument Response}
	\label{sec:pr_as_response}

With power recycling, the response of the interferometer to time-varying path length displacements is complicated by
the storage time of the recycling cavity, which imposes a bandwidth limit of approximately 350~Hz on the cavity
response. At frequencies ${\ll}350$~Hz, the response at the anti-symmetric port to arm length displacements reflects both a change in the Michelson fringe offset and in the cavity storage power. However, at frequencies ${\gg}350$~Hz, arm length displacements occur on a shorter time scale than the cavity can respond. In this limit, the power-recycled interferometer responds equivalently to a single-pass Michelson interferometer of the same optical power.

\begin{figure}[!tp]
	\centering
    \includegraphics[width=1\linewidth]{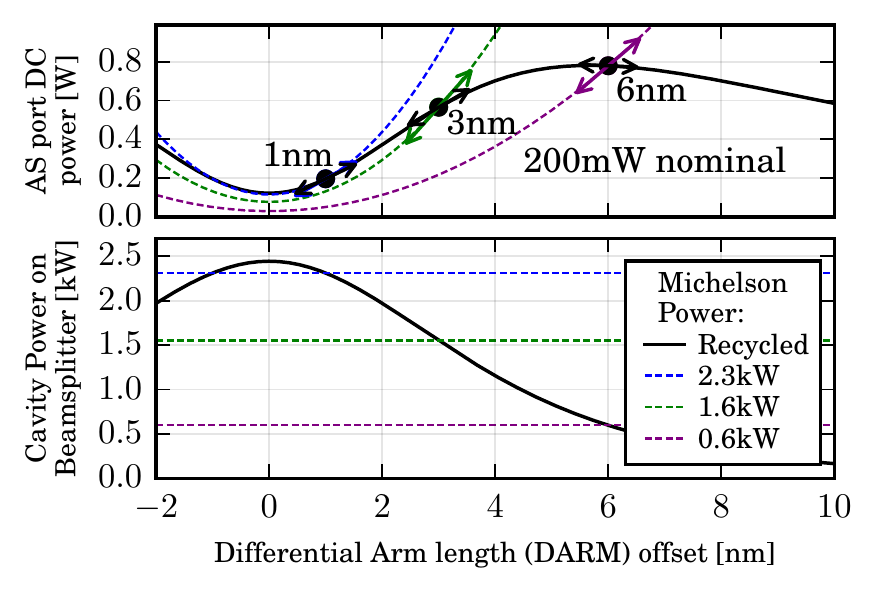}
    \includegraphics[width=1\linewidth]{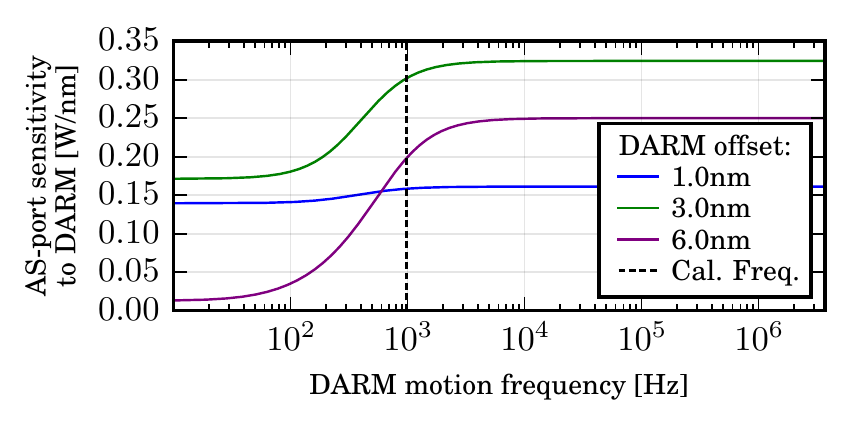}
	\caption[Transfer function of the power-recycled interferometers, shown at several optical path difference operating offsets.]{Instrument response of the power recycled Michelsons shown at several optical path difference operating offsets. A transition between the AC and DC asymptotic limits occurs at the cavity bandwidth of $350$~Hz. At frequencies below $350$~Hz, the response reflects changes in both the interference offset and the cavity power. At frequencies above the cavity bandwidth, the power-recycled interferometer behaves as a single-pass interferometer of equivalent power. The Holometer operates firmly in the AC asymptotic regime of the 1-nm offset curve.}
	\label{fig:pr_asport_response}
\end{figure}

Fig.~\ref{fig:pr_asport_response} shows the numerically-calculated transfer function of the power-recycled interferometers at several fringe offsets. The Holometer operates at an offset of approximately 1~nm, and all science and calibration signals are measured at frequencies $\ge1$~kHz. At this offset, the deviation from a single-pass Michelson response is $\le 2\%$ above $1$~kHz. The third plot, indicating the optical sensitivity, shows this difference at the calibration line frequency from the asymptotic response, which is small for the 1nm offset. Thus, for calibration purposes (see \S\ref{sec:indirect_calibration}) the instrument can be modeled as an equivalent high-power, single-pass interferometer. Neglecting the cavity correction underestimates the instrumental sensitivity. The degradation in sensitivity would be relevant for smaller interferometers with higher bandwidth recycling. For instance in tabletop versions of the experiment, the calibration and sensitivity would be affected over a much larger bandwidth. 

\subsection{Physical Environment}

The Holometer is situated in an old meson tunnel of the fixed-target beamline at Fermi National Accelerator Laboratory (FNAL). These tunnels are constructed of pre-built reinforced concrete slabs serving as the floor, set on a gravel bed at ground level. The slabs are 150~mm thick except at their edge, where they are 600~mm thick. Each slab is 3.7~m long and 6~m wide. The slabs are laid end-to-end with a small gap between them along the length of the tunnel. The walls are pre-formed U-shaped concrete sections 2.4~m wide placed on top of the floor sections. The entire structure is buried under a mound of gravel and dirt with a minimum thickness of 2.5~m, which formerly acted as a radiation shield. The tunnel structure was put in place more than 30 years ago and so is now a very stable platform. For the interferometer arm perpendicular to the tunnel, a newly constructed 1.9~m by 1.9~m square reinforced concrete slab 450~mm thick was built at ground level. The slab rests on three reinforced concrete pillars 600~mm in diameter and extending 1.8~m into the ground. A climate controlled enclosure constructed over the slab protects the electronics and end station vessels. All the vessels holding interferometer optics are bolted directly to the concrete slabs of the floor below the system.

\subsection{Optical Elements}

The following sections detail the laser optics chosen to achieve large resonant enhancement, as well as the infrastructure implemented to monitor and control the optical state.

\subsubsection{Laser}
Each interferometer uses its own 2W 1064nm laser (shown in Fig. \ref{fig:L_laser_launch}). These are Mephisto branded non-planar ring oscillators in Nd:YAG by Innolight.

\subsubsection{Injection Optics}
	\label{sec:injection_optics}

Prior to interferometer injection, the laser beam is conditioned to match the mode of the cavity at the power-recycling mirror. Each interferometer has a table of optics dedicated to beam preparation and to locking the laser frequency to the cavity using a Pound-Drever-Hall discriminant\cite{Black2001} (PDH, see \S\ref{sec:carm_loop}). Fig. \ref{fig:L_laser_launch} shows the principal optical components of the laser launch design. Immediately before injection, each table diverts approximately $1\%$ of its prepared beam power onto low-power detectors continuously monitoring the amplitude noise of the laser. These detectors may be operated as a Mach-Zehnder interferometer with a 0.6~m arm length imbalance to measure phase noise and calibrate against the PDH phase signal.

From the table, each injection beam is relayed to the input window of its central vacuum vessel by a set of three flat mirrors mounted to the wall, through 75-mm-diameter aluminum pipes in air. All of the mirrors are manually adjustable, and the last of the three is remotely controlled by two-axis, open-loop Newport piezoelectric stepper actuators (picomotors). Additionally the two steering mirrors, SM1 and SM2, on each table's output telescope are picomotor actuated for fine adjustments of the injection beam position and angle onto the optical axis of the interferometer recycling cavity.

\autofiguresvg{
  folder=figs/diagrams/,
  file=L_laser_launch, 
  caption = {Laser launch table for preparation of the injection beam. These optics perform three key functions: mode matching to the cavity (lenses not shown), locking the laser frequency to the cavity (see \S\ref{sec:carm_loop}), and continuously monitoring the phase and amplitude noise of the laser.},
}

\subsubsection{Power Recycling Mirror}

A partially transmissive 2-inch-diameter by 1/2-inch-thick power recycling mirror (PRM) is placed at the symmetric port
of each interferometer. The PRMs are fabricated from Corning 7980 0A low-inclusion fused-silica substrates and polished to sub-nm flatness by Coastline Optics.

The PRMs are coated by Advanced Thin Films to 80\% of their diameter by ion beam sputtering. Spectrophotometric measurements at 1064~nm found the transmission on the reflective surface to be 985.8~ppm and the reflection on the anti-reflective surface to be 13~ppm. Because the transmission through the PRM is the dominant loss in the power-recycled interferometer, the instrument is operated as a slightly over-coupled optical cavity.

Each PRM is mounted on a picomotor-actuated tip-tilt stage. These actuators make slow adjustments used for aligning each cavity into a resonant configuration. In each interferometer, the PRM and beamsplitter mounts each rest on a common critically-damped two-stage vibration isolation platform with a resonant frequency of 10~Hz. This platform consists of two stages of stainless steel plates, the bottom one weighing 27.8 kg and the top 18.9 kg. The stages rest on 19 mm diameter Viton balls, six beneath the lower stage and four between stages, constrained inside recessed holes in the lower platform.

\subsubsection{Beamsplitter}
	
The beamsplitters are 3-inch-diameter by 1/2-inch-thick optics, polished to sub-nm flatness by Coastline Optics, with 1.5~mrad of wedge. At $45^\circ$ incidence, the beamsplitters intercept the Gaussian beam with an acceptance of more than five times the waist size. They are fabricated from high-purity, low-OH$^-$-content Heraeus Suprasil 3001 substrate chosen for its low power absorptivity at 1064nm of $0.3\pm 0.2$~ppm/cm \cite{Loriette:2003}, which reduces its susceptibility to thermal lensing effects \cite{Winkler:1991}. With 2~kW of storage power in the power recycling cavity, the contrast defect due to thermal lensing in this substrate is expected to be only $10\pm 4$~mW.

The beamsplitters are coated by Advanced Thin Films to 80\% of their diameter by ion beam sputtering. The reflective surface has been measured by the coating vendor to have 49.991\% transmission. The back surface is anti-reflection coated to 60~ppm. The absorption due to impurities in the coatings has been measured via photo-thermal common-path interferometry by Stanford Photo-Thermal Solutions. The absorption on the reflective coating was found to be 0.95~ppm uniformly across the surface of a witness sample, and the absorption on the anti-reflective coating was found to be 1.3~ppm. Like the PRM, each beamsplitter is mounted on a picomotor-actuated stage, providing digitally-controlled dual-axis steering for slow alignment adjustments. 

\subsubsection{End Mirrors}
	\label{sec:end_mirrors}

The end mirrors are 2-inch-diameter by 1/2-inch-thick optics fabricated from Corning 7980 0A low-inclusion fused-silica substrates by Gooch \& Housego. They are coated to 80\% of their diameter by ion beam sputtering and have an effective radius of four times the beam size. This large mirror size is chosen to prevent losses in the tails of the Gaussian beams. Using an optical chopper and a lock-in amplifier, the mirror transmission has been measured to be sub-ppm. The back surface is anti-reflection coated in order to allow the instantaneous cavity power to be monitored via the light leakage through the mirror.

In order to select optimally matched pairs of end mirrors, the mirror surface structures were analyzed using both Zernike and Laguerre-Gauss decomposition of metrology data from Zygo interferometry. The actual mirror maps were also used in FINESSE Monte Carlo simulations \cite{Freise:2004} to determine the fraction of light scattered into higher order modes by mirror surface imperfections, which contributes to the contrast defect power. Pairs of mirrors were selected on the basis of matching in radius of curvature to $<10$~cm and with all other surface deviations predicted to produce a contrast defect of $<20$~ppm.

Each end mirror is clamped to an aluminum support ring that is driven by three Noliac SCMAP05-12mm piezoelectric transducers (PZTs). The PZTs provide three degrees of actuation over a total range of approximately 10~$\mu$m. These actuators are managed by the interferometer control system, and their performance is characterized in \S\ref{sec:pzt_drive}. The mirror is held by three clamps spaced $120^\circ$ apart. To minimize static mirror distortion from the mount clamping, a pad of indium, 5~mm in diameter and 20~$\mu$m thick, cushions each point of contact on both sides of the mirror.

\begin{figure}[!tp]
	\centering
    \includegraphics[width=1\linewidth]{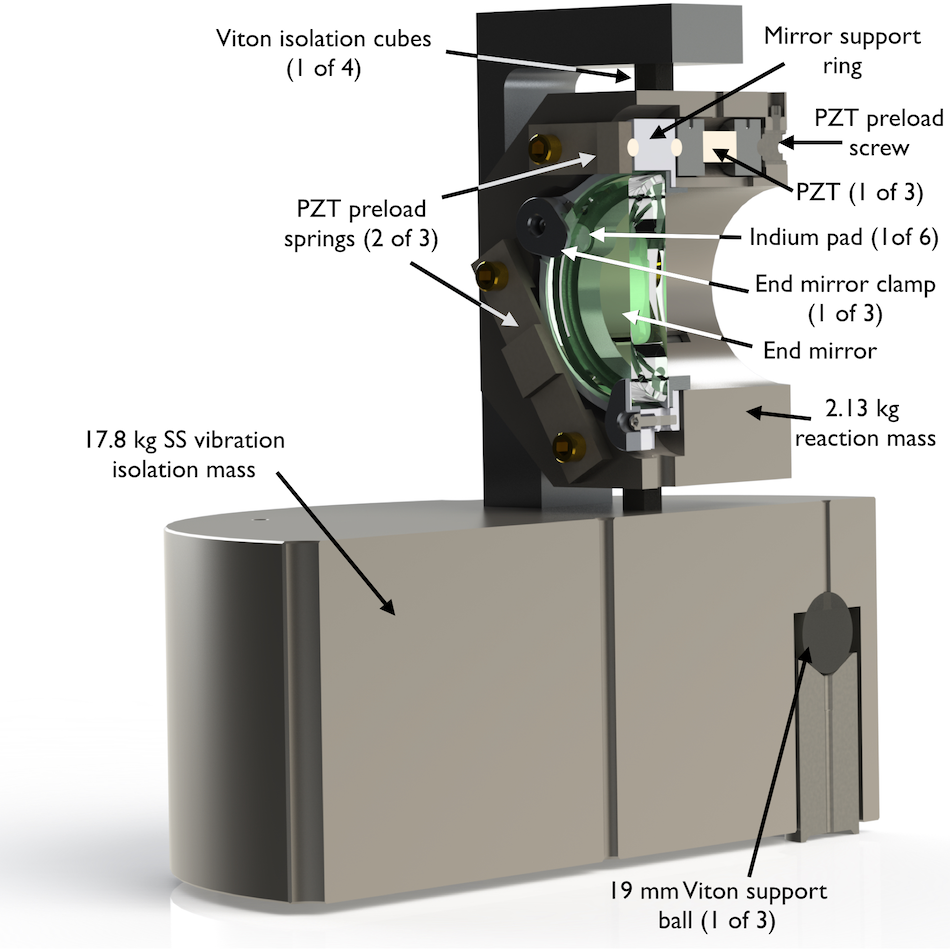}
	\caption[Sectioned view of the end mirror vibration isolation mount and PZT actuator.]{Sectioned view of the end mirror vibration isolation mount and PZT actuator. At the bottom is a mass mounted on three Viton balls. The plane of the balls contains the center of mass of the isolated system to decouple rotational from translational motion. The end mirror is actuated with three PZT actuators spaced at 120$^\circ$ around the mirror. The PZTs push against a 2-kg reaction mass which is isolated from the rest of the mount by Viton spacers. The PZTs are preloaded with springs. The mirror itself is held into a support ring by being clamped between six Indium pads, three on each side to minimize distortion of the mirror from the mount. The lowest resonant frequency of the drive system is above 1.5~kHz.}
	\label{fig:end_mirror_pzt_assembly}
\end{figure}

Each end mirror-PZT assembly is connected to a 2-kg stainless steel reaction mass, which is itself connected to a supporting mount. The mount has two stages of vibration isolation. The first stage, installed only in the two end stations located in the tunnel, where floor vibrations up to 100~Hz are present, isolates the entire mirror-mount structure from the ground with a resonance at 11~Hz. It consists of an 18-kg mass mounted on three Viton balls, whose plane contains the center of mass of the isolated system to decouple rotational and translational motion. This mount is critically coupled while the outside of the vessels are maintained at 37~C by a linear proportional thermostat system as the damping and stiffness of the Viton is temperature dependent. The thermal stabilization also reduces slow drift of the alignment. The second stage of the mount isolates the mirror and the reaction mass from the support mount, so that PZT drive motion pushes purely on the reaction mass. This is done to avoid exciting mechanical resonances of the rest of the mount. Fig.~\ref{fig:end_mirror_pzt_assembly} shows this design.

\subsubsection{End Mirror Transmission Optics}

The light transmitted through each end mirror is split by a beamsplitter and directed onto a photodiode and a video camera. Each optical set-up is mounted on a 12-inch by 12-inch breadboard bolted directly to the back of its end station cube. The photodiode signal provides a control-signal proxy for the cavity storage power, and the video camera is used to view the position and mode quality of the beam inside the cavity. Manual irises placed between the end mirrors and transmission cameras are used to center the beams on the end mirrors during alignment procedures.

\subsubsection{Output Optics}
	
The 3.6-mm interference beam exiting each interferometer at its anti-symmetric port is telescopically focused down to 1-mm size and directed onto a set of high-frequency signal detectors, low-frequency control system detectors, and a video feed. Each optical set-up is mounted on a 12-inch by 24-inch breadboard supported by 8-inch-diameter tubes bolted directly to the floor. Of the approximately 200~mW of power in the interfering beam, 1\% is picked off for control and monitoring purposes. The remaining 99\% is directed onto two New Focus 1811 detectors which have been modified to absorb high power (see \S\ref{sec:modified_detectors}). These detectors are protected by a digitally-controlled shutter that is triggered to open when the exiting power is at a safe level for the detectors.

The low-power control pickoff is split by a beamsplitter and directed onto a New Focus 2903 quadrant photodetector (QPD), which provides the anti-symmetric port control signals, and a video camera, used for monitoring the mode quality and angular alignment of the interfering beam. As the QPD has an adjustable gain, it is used during power-recycling mode as well as during low-power operations, where the PRM is misaligned to make a 1~mW single-pass Michelson. The quadrant detector's 2-mm total collecting area is divided into four equal-area sections. The sum of the entire receiving area is used to for the DC-readout control signal, while the differences between the quadrants are used as error signals for the angular alignment control loop (see \S\ref{sec:alignment_loop}). The sum channel was modified, removing a stock output low pass filter to provide instead a flat response to at least 1~kHz.

\subsubsection{Baffles}

To prevent multi-path interference from light reflected and scattered by the inner walls of the vacuum tubes, a set of conical baffles is installed in each of the interferometer arms, spaced such that neither beamsplitter nor end mirror has a direct line of sight to the tube wall. Each baffle is a truncated cone (half-angle $35^\circ$, or 0.61~rad) with an outside diameter of 150~mm, matching the inner diameter of the vacuum tube, and an inside diameter of 110~mm. The baffles are constructed from sheets of thin stainless steel spot-welded into a cone that fits snugly into the vacuum tubes. Prior to assembling the 40-m arms, the baffles were pushed into the individual 10-foot tubes using a long pole. Each arm contains a total of 31 baffles spaced in a geometrical series derived from the design in the LIGO interferometers\cite{flanagan:1994}. Starting from each end of the arm and working inwards, the first 30 baffles are positioned a distance $l_n=\alpha^n\,l_0,\; n=0,1,\dots,14$, from each end, where $\alpha=1.28$ and $l_0=0.5$~m. The final baffle is positioned at the center of the arm.

\section{Vacuum System}

The PRM, beamsplitter, and end mirrors of each interferometer are housed under vacuum. This is done to reduce phase
noise caused by fluctuations in the index of refraction of air, which would otherwise dominate the noise budget, as well
as to avoid burning hydrocarbons onto the optical surfaces exposed to high power. Separate, identical vacuum systems are
constructed for each interferometer. The following sections describe their design and implementation.

\subsection{Partial Pressure Requirements}

Based on previous experience from the construction and operation of the LIGO vacuum system, keeping the optics free of hydrocarbon deposits imposes the most stringent requirement on the Holometer vacuum systems. Every hydrocarbon deposit of 0.1 monolayer thickness on the surface of the beamsplitter contributes 2~ppm of absorption losses \cite{Winkler:1991}. Accordingly, the Holometer vacuum system is designed to minimize the rate of hydrocarbon deposition on optical surfaces, which is estimated as follows.  The rate of deposition is given by
\begin{align}
r \; \textrm{(molecules/cm)}^2\textrm{/s} = S_{\rm c} \cdot \rho \cdot v_{\rm th} \;,
\end{align}
where $S_{\rm c}$ (sticking coefficient) is the fraction of molecules striking the surface which stick, $\rho$ is the density of molecules ($3.5 \cdot 10^4/\textrm{cm}^3$ at $10^{-10}$~Pa), and $v_{\rm th}$ is the thermal velocity of the parent hydrocarbon ($3 \cdot 10^4$~cm/s at room temperature). This yields a deposition rate of $r = 3.5 \cdot 10^7/\textrm{cm}^2/\textrm{s}$. One monolayer is $3.5 \cdot 10^{14}/\textrm{cm}^2$, so, at a partial pressure of $10^{-10}$~Pa, 0.1 monolayers builds up in $10^6$ seconds, about ten days.

Using the partial pressure of hydrocarbons after baking as an acceptance criteria during construction is not practical,
since testing each part with a full bake and vacuum system takes more labor than assembling the final system. Therefore,
a test has been adopted which measures the number of monolayers of hydrocarbons on the surface of each fabricated part
prior to baking. To estimate the surface contamination, high-performance liquid chromatography (HPLC) grade isopropyl
alcohol (IPA) is poured inside each part and the concentration of absorption at the wavelength of C-H bonds is measured.
Experience with previous vacuum systems motivated setting an acceptance threshold of $<5$~monolayers before baking, as
the bake typically reduces the surface contamination by at least two orders of magnitude. All of the parts passed this
acceptance criteria. After construction and baking of the entire assembled system, each vacuum system has achieved
hydrocarbon partial pressures $< 10^{-10}$~Pa. At this level, no increase of optical absorption losses has been
detected over time.

\subsection{Arm Tubes}

Each 40-m interferometer arm is constructed of 12 sections of 10-foot-long stainless steel tube with 8-inch ConFlat flanges at the ends. The stainless steel tubes are supported on rollers every 10~feet. A 100-l/s ion pump is installed at the end of each arm via a reducing tee with a 2~3/4-inch ConFlat flange. Another reducing tee with a 6-inch ConFlat flange and 4-inch gate valve is installed at the center of each arm, where a 4-inch turbo pump can be attached to bake out the tube. The arm tubing is covered with two layers of fiberglass pipe insulation, separated by a layer of aluminized mylar, for a total thickness of 5~inches. The insulation allows the arm tube to heat to a temperature of 150~C during baking. After assembly, each arm was baked for a minimum of three days by applying 100~A of current through the tube, with the turbo pump operating at the center.

\subsection{Central Vessels}

The central vacuum vessels, each housing the beamsplitter and power recycling mirror of one interferometer, are custom-built stainless steel cylinders 600~mm in diameter and 330~mm in height, with a bottom plate 25~mm thick. Each vessel is closed off with a rubber o-ring-sealed top plate, also 25~mm thick. The positions of the central vacuum vessels are not adjustable. Each vessel stands on three aluminum tubes 200~mm in diameter and 13~mm thick. Under each tube is a 150-mm by 150-mm by 6-mm aluminum support plate, which was positioned on the somewhat uneven floor with a thick layer of epoxy underneath. The tubes and vacuum vessel were then placed on top and the epoxy allowed to harden. Three dog clamps drilled into the concrete floor push down on the bottom plate of the vacuum vessel near the support tube underneath. The lowest horizontal resonance of the mount is calculated to be $>200$~Hz. 

\subsection{End Station Vessels}

The end station vacuum vessels, each housing an end mirror, are 10-inch 6-way stainless steel ConFlat cubes purchased from Lesker. A custom-built bottom plate mounted to the bottom ConFlat flange of the cube extends outside the cube itself and serves as its support base. The cubes stand on three aluminum tubes similar to those of the central vessel, but each of these additionally has a top plate with a 2-mm boss in the center. When the cube is placed on top of the cylinders, these bosses act as compliant elements. As with the central vessel, the cubes are secured to the floor with dog clamps drilled into the concrete. In this case, however, the compliance of the supporting tubes permits small mechanical adjustments of the pitch and roll orientations of the end station cube. 

\begin{figure}[!tp]
	\centering
    \includegraphics[width=1\linewidth]{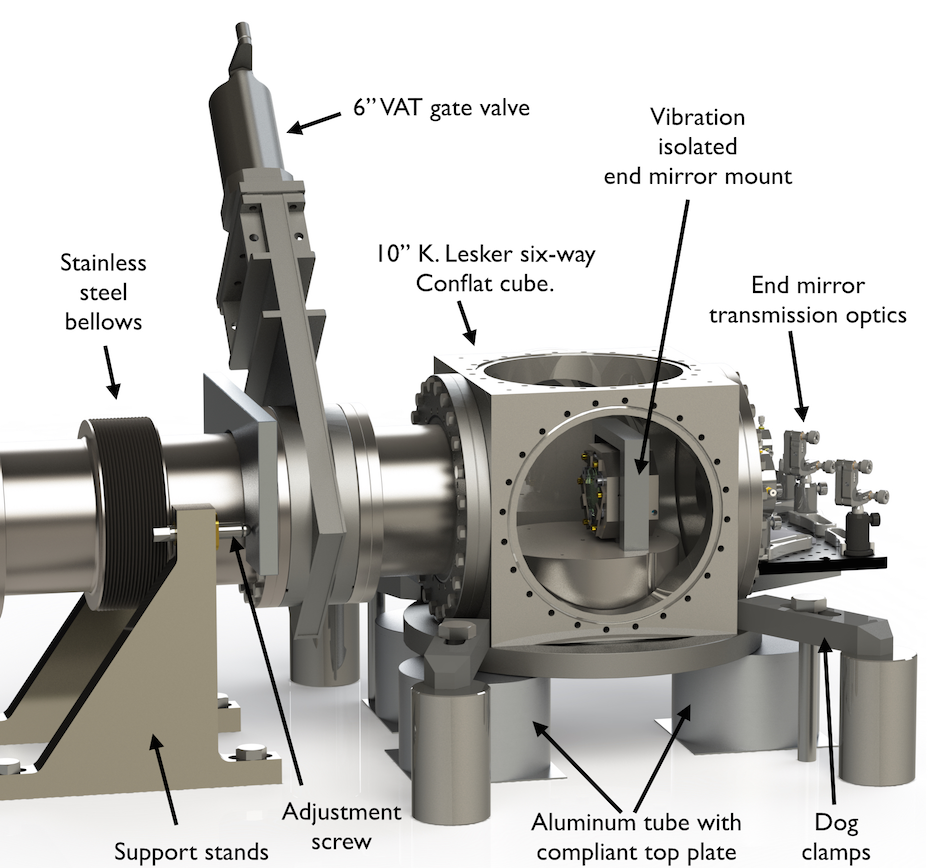}
    \caption[End station assembly with top and side flanges removed.]{
      End station assembly with top and side flanges removed. The end station cube is mounted on three compliant tube mounts and clamped to the floor with dog clamps. The compliance permits adjustment of the pitch angle. The two support stands with adjustment screw permits changing the length of the arm and the yaw angle of the mount. With these adjustments, the end station angle can be changed to bring the mirror within range of the PZT actuators on the end mirror. The bellows mechanically isolate the end station from the vacuum tubes. The end station can be vented for servicing without venting the tube or detensioning the bellows. The transmission optics are mounted on the back ConFlat flange, which has a 2-inch window.
    }
	\label{fig:endstation_assembly}
\end{figure}

The end station cubes are mechanically isolated from the arm tubes by stainless steel bellows, whose length can change
by up to 90~mm to accommodate thermal expansion. The cubes are supported against air pressure on the evacuated bellows
by two stainless steel stands bolted to the floor, one on each side of the tube. Each support stand is outfitted with a
5/8-inch 80-threads/inch adjustment screw which contacts the arm tube flange at the center height of the tube. These
screws provide for mechanical adjustment of the end station position and orientation. Adjusting both screws in common
lengthens or shortens the bellows, and thus the arm length and pitch of the end mirror, whereas adjusting them differentially rotates the yaw-axis of the end mirror. A 150-mm diameter gate valve, installed between the bellows and cube, permits venting the end station vessel for access to the optics without venting the tube or releasing the tension on the bellows. Fig.~\ref{fig:endstation_assembly} shows the end station design. 

\section{Control System}

Each power-recycled interferometer requires a control system to maintain stable linear operation. This system maintains the optical resonance conditions that allow the instrument to both store high power and achieve the appropriate detuning from perfect destructive interference at the anti-symmetric (dark) port. The following sections describe the hardware and software implementation for these controls. The final section then discusses each optical resonance condition and the feedback control servo enforcing it.

\subsection{Piezoelectric End Mirror Actuation}
	\label{sec:pzt_drive}

Each end mirror is actuated with three piezoelectric (PZT) drivers capable of moving 10 $\mu$m in the direction perpendicular to the reflective surface, as described in \S\ref{sec:end_mirrors}. The control system of each interferometer engages these actuators to maintain both the longitudinal operating offset and the angular alignment of the instrument. The first PZT resonance occurs at 1.5~kHz, limiting the stable operation bandwidth in spite of compensating notch filters. These resonances are regularly re-measured, as they are observed to drift up to $10\%$ in frequency over many weeks. This drift has been associated with both hysteresis over the 10$\mu m$ range as well as with operating temperature. The PZT's have a capacitance of 2$\mu$F and are driven by a linear high voltage amplifier with an output impedance of 1k$\Omega$. The RC low pass filter is compensated in the digital control system.

The mechanical resonances are measured by injecting white noise into the PZT drive of one end mirror while the differential-mode loop (see \S\ref{sec:darm_loop}) is engaged. The transfer function of drive signal into physical motion of the mirror is inferred from the optical response measured at the dark port. Individually in each arm, digital notch filters are applied within the differential-mode servo loop to eliminate the gain spikes at these resonant frequencies. The added stability allows the servo bandwidth to be increased. Because of the sensitivity of the PZT resonance locations and the excitation of only a single end mirror, the closed loop gain is not corrected for these measurements, but they are iterated as the loop stability improves. Rather than exact matching, the resonance center frequencies and Q-factors are estimated and pairs of notches are added to engulf the resonance, accounting for some later drift. Adding pairs of widened notches adds robustness at the cost of additional phase delay reaching down to the 600Hz unity-gain frequency (UGF).

\subsection{Low-Speed Digitization Electronics}

Two National Instruments (NI) PXI-7852R units are employed to monitor, condition and generate signals for each interferometer. Signals for the feedback loop control (see \S\ref{sec:control_loops}) are routed exclusively through one unit, while auxiliary signals are routed through the other. Each unit consists of an FPGA controller interfaced to an 8-channel analog-to-digital converter (ADC) and an 8-channel digital-to-analog converter (DAC), both of which can sample up to 750~kHz at $\pm10$~V over 16~bits. All of the I/O channels are conditioned in a custom filter buffer board which provides anti-alias filtering of the inputs and DAC quantization filtering of the outputs. In both cases, this conditioning is implemented through analog 40-kHz, 2-pole Butterworth filters. 

The input voltage range of each channel is individually adjustable through an AD8253 programmable-gain instrumentation amplifier in the filter buffer board. These analog gains are set using the digital outputs of the PXI-7852R, which powers the filter buffer board via a DC-DC converter running from a 1-A, 5-V line. The input signals can be amplified to ranges of $\pm0.005$~V through $\pm5$~V and the output signals can be attenuated to ranges of $\pm0.01$~V through $\pm10$~V. These adjustments allow convenient tradeoffs between dynamic range and bit-noise. Each amplifier has an input voltage noise of 10~$\rm nV/\sqrt{\rm Hz}$ near 1 Hz, with the exception of the gain~1 setting, where the input-referred noise is 45~$\rm nV/\sqrt{\rm Hz}$. The analog inputs are differential with common-mode rejection of over 60~dB below 40~kHz.

\subsection{Controls Software}

The FPGA running the controls software uses a micro-controller architecture, where an instruction pipeline feeds the eight input signals through stages of matrix elements and cascaded filter banks, represented through an adjustable sequence of control codes. The pipeline runs 1400 digital bi-quadratic filters (BQF) at 32~kHz. Before entering the pipeline, raw input signals to the 16-bit ADC are sampled at 128~kHz and immediately fed into a single BQF for digital anti-aliasing, then downsampled to 32~kHz. The 32-kHz signals are routed through matrix elements with configurable coefficients, multipliers (i.e. mixers), and dividers to synthesize discriminants and linearize the interferometer signals, to demodulate dithers, and to transform bases between detector, controller, and actuator degrees of freedom. BQF filter banks are applied to whiten signals, impose control policies, and invert actuator responses. Final actuator signals are then upsampled back to 128~kHz and fed through a smoothing BQF filter and output to the 16-bit DAC. Internally, the signals are 32-bit fixed-point numbers, however the direct-form-1 BQF filters increase to 64bit within the feedback portion, preventing increased quantization noise and design errors in the filters. 

A python management script sets up the topology and interfacing between the interferometer state variables and the settings of the FPGA micro-control code. The interferometer state variables are hosted by Experimental Physics and Industrial Control System (EPICS) software \citep{epics}. This state is manipulated via the Motif Editor and Display Manager (MEDM) software, providing a graphical interface for users to monitor interferometer operation and adjust control parameters in real time \citep{medm}. 

Automated lock-acquisition scripts provide high-level interferometer state management. These scripts interface through
EPICS to repeat action sequences for the low-level control parameters. The configurable parameters of the controls
scripts are themselves hosted as EPICS variables, allowing a consistent storage and manipulation interface for low and
high level settings. The lock sequence monitors the interferometer light levels, sequencing when signals fall outside of
nominal high power operations. The auto-lock software additionally triggers the data acquisition system to prevent
excessive storage during interferometer dead-time.

\subsection{Feedback Control Loops}
	\label{sec:control_loops}

In the following sections, the design motivations and optimizations for the feedback control loops are described. These are separated into the common (average) and differential basis for the arm lengths. The common arm length (CARM) affects the resonant laser frequency, while the differential (DARM) length sets the interference at the beamsplitter. Both must work in tandem as the Michelson interference affects the effective Fabry-Perot cavity and PDH error signal bandwidth, and the recycled power storage affects the linearity and gain of the DARM servo error signal.
	
\subsubsection{Common-Mode Loop}
\label{sec:carm_loop}

The laser frequency is used to actuate the common-mode loop on to a multiple of the 3.83~MHz free spectral range of the
recycling cavity. Each interferometer uses an electro-optic-modulator (EOM) driving with an index of $.01$~rad outside
of the cavity bandwidth, one at 20.6~MHz and one at 24.49~MHz. These are well above desired data-acquisition measurement band as well. The reflection ports use photodiodes (RF PD in Fig. \ref{fig:L_laser_launch}) customized similarly to the AS-port high power diodes, except with an additional inductor pair forming a tank-circuit around the diode. The pair of inductors is folded to reduce external magnetic pickup.

The PDH error signal is linear only within the bandwidth of the cavity, which changes dramatically as the loss through the AS port moves to destructive interference. Furthermore, the error signal gain is proportional to field amplitude of the carrier and sidebands, making it scale proportionally to the power gain or the finesse, $\mathcal{F}$. The bandwidth scales inversely to the field gain, making it proportional to $1/\sqrt{\mathcal{F}}$.

The common mode loop is electrically isolated for noise. This servo actuates through the fast frequency adjustment piezo on the Mephisto laser. In suppressing phase noise between the interferometer cavity and laser, it can potentially imprint RF environmental noise detectable in the two-interferometer cross correlation. Regardless of the design, some correlated phase noise is imprinted by the servos, and measured bounds to characterize this instrumental background are essential (see \S \ref{sec:pm-mitigation}).

For both speed and isolation, the CARM servo is implemented in analog with two Stanford Research SR560 preamplifiers, powered from the same AC line as the laser. Additional passive shaping is applied between the output of one amplifer and the input of the other, and the second amplifer has additional LC notches and RC filters above 100kHz to compensate for the laser PZT resonances and to reduce the gain of environmental RF pickup. The CARM loop is unconditionally stable at any finesse, even as the PDH gain and cavity pole change through the DARM servo lock acquisition.

\begin{figure}[!tp]
	\centering
    \includegraphics[width=1\linewidth]{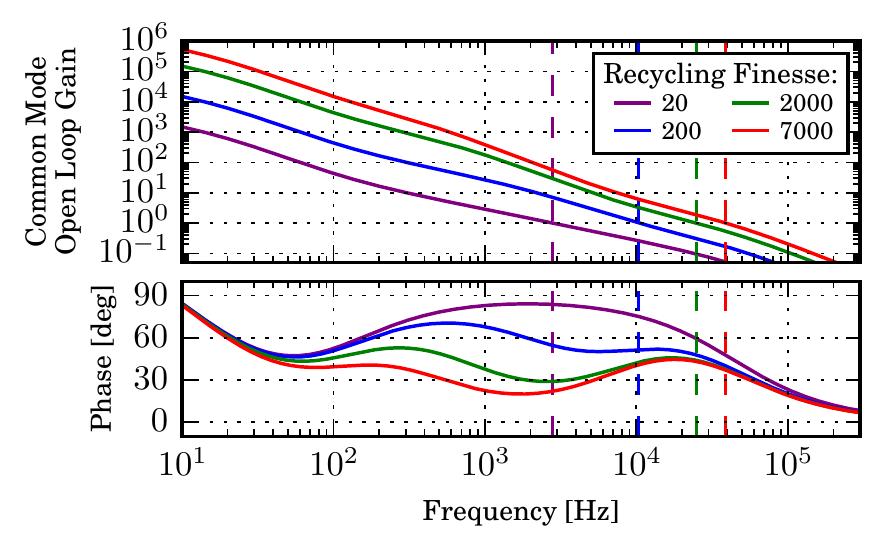}
    \caption{
      Open-loop gain of the cavity frequency loop, through changing optical response. As the finesse rises, the gain rises proportionally, but the cavity pole frequency is reduced. The dashed lines show the rising unity-gain frequency, while the phase shows the increasing storage time. The response is designed to ensure stability over the range of finesse.
    }
	\label{fig:compensated_CARM_loop}
\end{figure}

Fig. \ref{fig:compensated_CARM_loop} shows the employed loop design accounting for the changing optical response. The
design follows from a 25kHz unity-gain frequency (UGF) at the nominal finesse. An additional zero is placed to
compensate for the cavity-pole bandwidth decreasing towards 350Hz in the discriminant. This zero could make the servo
unstable at low finesse, but the UGF starts below this zero and raises with the optical gain towards the designed zero
as the cavity pole moves to cancel it. The cancellation is not total, boosting the loop gain and noise suppression. An additional lag-lead filter adds further gain at low frequencies. This allows a DC gain $>1000$ even with a finesse of 20, so that the common mode loop readily converges. This provides reliable cavity gain and therefore reliable error signals to the differential servo so that its nonlinear response can be linearized around predictable offsets.

\subsubsection{Differential-Mode Loop}
	\label{sec:darm_loop}

To maintain a constant DC output power at the dark port, the optical path difference, DARM, denoted as $\delta\hspace{-.1em}L$, must be held at a constant operating offset. The error signal to determine this path difference is linearized around offsets of the black signal traces shown by the two top graphs of Fig. \ref{fig:pr_asport_response}. The AS-port power and the cavity power are both used as discriminants for DARM. Furthermore, these signals are only well-defined while the CARM servo is enforcing the optical power gain from resonance.

The lock is acquired first using the cavity power as measured by transmission through the North end mirror. 10W is the initial operating point, corresponding roughly to a finesse of 20, where the CARM loop can first acquire. Here, DARM must be within $\pm10$nm of the destructive interference fringe. The ratio of the static input power over the measured cavity power is used as a DARM discriminant of the finesse rather than the cavity power itself. Before the loop is fully shaped, the cavity power is not linear over the residual noise and biases towards cavity power lower than the offset. The synthesized ratio signal is also nonlinear but is biased towards higher power. An offset is applied to the ratio signal for an initial lock point at the finesse of 20. The initial loop gain is tuned such that as the loop is closed, the offset causes an impulse that sweeps DARM through a full fringe, while being sufficiently slow to prevent overshoot in the CARM and DARM loops as the fringe offset crosses the appropriate point.

DARM motion conversion into an optical cavity power signal is low-passed by the pole of the storage cavity. This pole is digitally inverted at the FPGA. The offset of the synthetic signal implicitly determines the fringe detuning from destructive interference. A series of linear interpolation points transition the offsets and loop gains to stably increase the cavity power to 1kW - where the cavity is critically coupled. At critical coupling, the discriminant is smoothly transitioned to use the AS-port fringe power. The interference power output is first divided by the cavity power to synthesize the sinusoidal response of a non-recycled Michelson interferometer:
\begin{align}
  T\tb{AS} &= \frac{P\tb{AS}}{P\tb{BS}}
  \label{eq:T_AS}
\end{align}
This response, known as the AS-port transmissivity, is broadband (flat), unlike the optical response shown at the bottom of Fig. \ref{fig:pr_asport_response} and the cavity power signal. Using the ratio of the two measured powers, it is furthermore insensitive to drifting power levels. The error signal offsets are then transitioned to the nominal operating point of nearly 100PPM interferometer AS-port transmissivity, corresponding to roughly 100mW of fringe light and 100mW of contrast defect.

\begin{figure}[!tp]
	\centering
    \includegraphics[width=1\linewidth]{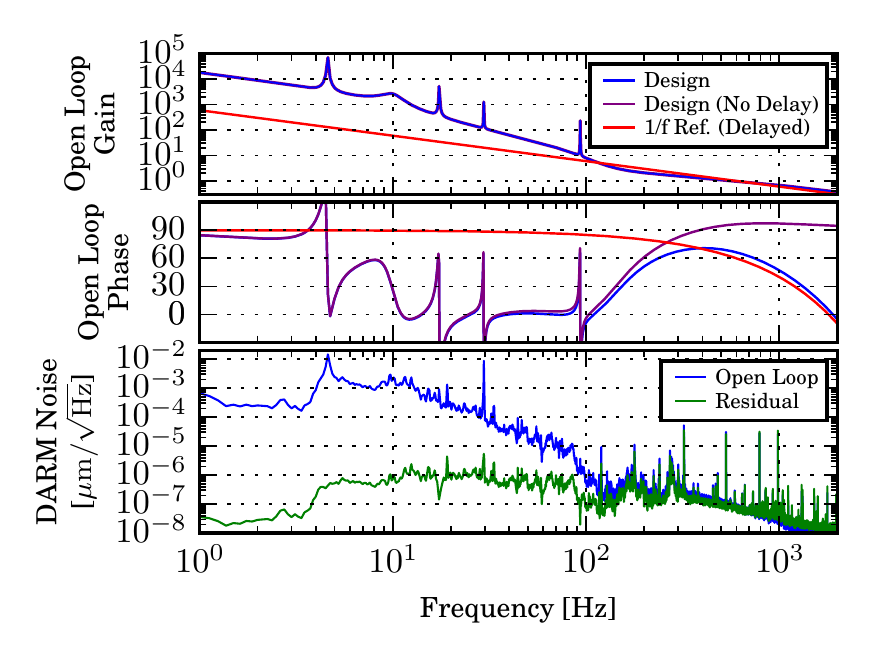}
    \caption{
      The top two plots show the designed open loop gain as compared to a 1/f reference. The sampling, Nyquist filter and PZT notch delays are included except on the purple trace. The resonant loop features are to reduce intermittent external sources. The sub 1/f response of the designed filter recovers phase rapidly despite the intrinsic delay of the notched system. The total residual RMS measured with and without including the 4 calibration lines is 30 and 10 pm/$\sqrt{\mathrm{Hz}}$, respectively.
  } 
	\label{fig:seismic_noise}
\end{figure}

The servo loop shape is shown in Fig. \ref{fig:seismic_noise}. Near the UGF, the loop is $1/f$, but is ``boosted'' to
have additional gain at low frequencies to flatten the $1/f^2$ seismic spectrum. The total delay of the sampling rate,
Nyquist filters and PZT notches causes a delay-like phase lag that is $50^\circ$ by 1kHz. The UGF is set to a nominal 600Hz, balancing between this overall gain, this delay, and pzt resonance stability. The boosted loop includes a complex pole-zero pair which causes a sub-$1/f$ response between 150-400 Hz. This causes a rapid phase recovery from the $1/f^2$ boosts and provides additional phase margin at the UGF. The purple trace shows the design without additional delay, and the red trace provides a comparison to a standard $1/f$ control loop.

\subsubsection{Angular Alignment Loop}
	\label{sec:alignment_loop}

The contrast defect is affected by the overlap of the two transverse field profiles of the end mirror return beams.  The total defect is the sum of the defect from higher-order modes, $\epsilon_\text{hom}$, and the defect due to imperfect overlap. Since the beamsplitter is near the 3.6mm waist radius at the PRM, the beam radius of curvature may be ignored, giving
\begin{align}
  \epsilon_\text{cd} = \epsilon_\text{hom} + \left( \frac{\Delta \theta \cdot 39\mathrm{m}}{3.6 \cdot 10^{-3}\mathrm{m}}\right)^2
\end{align}
where $\Delta \theta$ is the residual RMS angular displacement of the end mirrors.
To achieve an acceptably small fraction of the contrast defect from alignment, $<10$~ppm, $\Delta \theta$ must be suppressed to ${<}0.3~\mu\text{rad}$. 
The differential angular alignment of the beams is controlled to maintain the overlap. 

The error signal to control the differential alignment is provided by a New Focus 2903 QPD at the antisymmetric port
viewing the interference beam from a 1\% pickoff. Because of the small DARM operating offset, the QPD is sensitive to
the transverse field interference of the Hermite-Gauss $H_{01}$ mode of the misalignment with the $H_{00}$ mode of the
DC carrier. This is different from the usual of a QPD for beam position sensing and requires that the Michelson be
detuned from perfect destructive interference. The QPD does also sense the common return angle from the position of the
output beam, but whereas the differential angle signal is ejected from the Michelson immediately, the common angle
signal is reduced by the total transmissivity (field, not power) of the destructive interference, which is held by the
control system to be $t\tb{AS}^2 = {\sim}100$PPM. The vertical and horizontal set-points of the differential angle loop are implemented as offsets applied to the QPD signals.

Only the differential alignment is actively controlled. The injection alignment drifts, lowering the power coupling into the cavity from the typical maximum of 96\%. Furthermore, although the QPD is a factor $100$ more sensitive to differential than common angular fluctuations, the drift of the common angle slowly moves the optimal set-point for the differential loop. Periodically, the data system is disabled for manual reoptimization of the uncontrolled alignment degrees of freedom. During this process, an operator re-tunes the common-angle and injection alignment. New offsets are then loaded into the differential angle control loop to center the interference pattern in camera images. This procedure maximizes the online DARM sensitivity reported by the automated control system.

The differential angle servo is operated with a 100 Hz UGF. In one of the interferometers, the static mode of the contrast defect light alters the beat signal such that X- and Y-angle misalignments cause a degenerate QPD response. For this signal, the dithered $H_{00}$ mode at a calibration-line frequency (see \S\ref{sec:calibration_line}) is demodulated with the $H_{01}$ mode of the misalignment to generate a decoupled error signal, but with a much-reduced bandwidth of 20~Hz and a lowered SNR compared to the direct QPD signal.

\section{Data Acquisition System}
	\label{sec:correlator}

The output beam power is measured at the AS port of each interferometer by a set of custom-modified photodetectors,
whose signals are continuously sampled by the data acquisition system. Quantum-geometrical effects will appear as
correlations between the two interferometer signals on time scales shorter than the round-trip light-crossing time of
the apparatus, $2L/c=262$~ns (see \S\ref{sec:intro}). This imposes a minimum sampling frequency of $2 \cdot 3.82$~MHz.
To fully resolve the time scale of the effect, the Holometer samples at 50~MHz, more than 1000 times faster than do
gravitational wave detectors. This carries the advantage of sampling band where the contributions of seismic, thermal, and mechanical noise are negligibly small above 1~MHz. With a total data stream rate of~5.8 TB/hr, it also presents a unique set of technical challenges. The Holometer feeds the fast data streams into a novel high-speed data processing pipeline. It performs real-time spectral analysis on the 50~MHz-sampled interferometer and environmental monitoring signals, reducing the data storage requirements to $< 50$~GB/hr with no loss of statistical information. The following sections describe the design and implementation.

\subsection{High-Speed, High-Power Photodetectors}
	\label{sec:modified_detectors}

At the nominal operating point of approximately 100PPM AS-port transmissivity, each interferometer outputs roughly 100mW
of fringe light and 100mW of contrast defect. The total output power of approximately 200~mW is split by a beamsplitter
and directed onto two modified New Focus 1811 photodetectors, each of which is engineered to absorb 100~mW of DC power.
The photodetectors provide both a low-frequency amplification circuit for interferometer control signals and a
low-noise, high-frequency amplification circuit for radio-frequency signals for scientific analysis.
	
Each modified high-power detector contains a 2mm InGaAs photodiode from GPD Optoelectronics Corp. mounted in a TO-5
package. The photodiode is reverse-biased using a LM340 voltage regulator at 7~V. Low-value resistors are
used along the biasing chain, causing a drop in the bias voltage to 6~V at the nominal 100mA operating photocurrent.
These values are chosen because space charge-related capacitance becomes significant at bias voltages $<5$~V and breakdown of the photodiode occurs at a bias voltage of 12~V. To dissipate the 0.6~W of Joule heating, the TO-5 package is wrapped with layer of silicone-based thermal putty and thermally contacted with an aluminum heat sink.

The low-frequency channel uses an INA128 instrumentation amplifier to measure the voltage drop across a 5$\Omega$ resistor in series along the photodiode biasing circuit. It provides a gain of 10~V/A and a bandwidth of approximately 150~kHz, limited by the filtering circuitry along the photodiode biasing line. The $\rm 8\;nV/\sqrt{\rm Hz}$ noise of the instrumentation amplifier near 1~kHz is subdominant to the digitization noise of the control system.

The high-frequency channel uses a 7.2k$\Omega$ Philips NE5210D transimpedance preamplifier, AC-coupled to the photocurrent with a 500pF capacitor. The 100mA DC photocurrent is shunted to ground with a Vishay/Dale TJ31UEB270L 27$\mu$H toroidal inductor. Combined with the 60$\Omega$ input impedance of the preamplifier, the inductor and capacitor form a crossover filter centered near 1~MHz. This crossover frequency is chosen to suppress large-amplitude, acoustically-induced optical signals at several 100~kHz, which would otherwise saturate the preamplifier. This LC crossover filter has a resonant peak at 1.5Mhz. The lossy inductor core material acts as roughly 100 Ohms of resistance at the resonance frequencies to form an RLC with a Q of 2. The input pin of the amplifier is also shorted to ground with 10~pF to suppress unstable amplifier oscillations at several 100~MHz. This capacitance, combined with the 150pF capacitance of the reverse-biased photodiode, forms an additional low-pass filter with the 60$\Omega$ input impedance of the preamplifier, with a pole frequency near 10~MHz. The high-frequency channel is shot noise limited at a photocurrent of 0.5~mA, with an approximately flat amplifier noise level of $\rm 3.5\;pA/\sqrt{\rm Hz}$ from 1-10~MHz.

\subsection{High-Speed Digitization Electronics}
\label{sec:high_speed_adcs}
	
The high-frequency photodetector readouts and the environmental monitoring signals are routed through NI PXIe-5122 high-speed ADC units. These dual-channel ADCs can sample up to 100~MHz at $\pm10$~V over 14~bits. Each interferferometer has two dedicated ADC units. One unit receives the high-frequency readouts of the two photodetectors and the other receives two environmental monitoring signals. The input channels are anti-aliased via analog 20MHz, 2-pole Bessel filters. The input voltage range of the two channels on each unit is adjustable through a programmable-gain instrumentation amplifier. The input signals can be amplified to ranges of $\pm0.1$~V through $\pm5$~V, with a minimum input voltage noise of $\rm 2 \; nV/\sqrt{\rm Hz}$ at the highest gain setting.

\begin{figure}[!tp]
	\centering
    \includegraphics[width=0.48\textwidth]{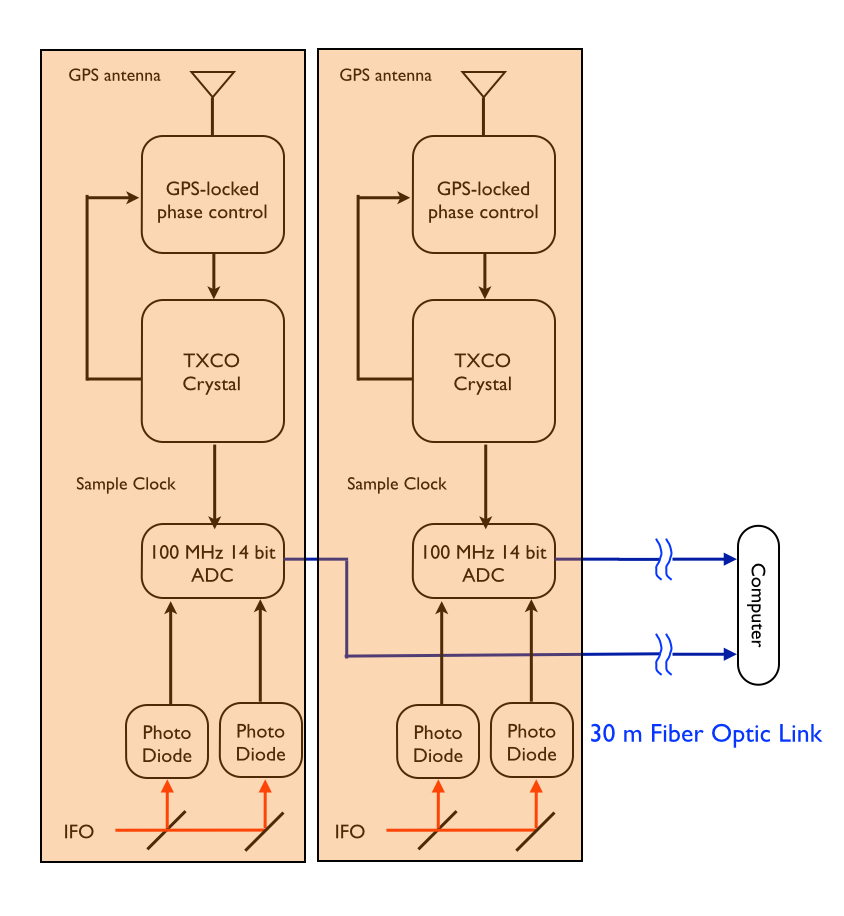}
	\caption[Schematic of the high-speed data acquisition system.]{Schematic of
the high-speed data acquisition system. Each electrically-isolated ADC unit is
synchronized to the others by phase-locking its sample clock to a common
GPS-provided clocking signal. For simplicity only two of the four PXI chassis
are pictured.}
	\label{fig:fast_adc_schematic}
\end{figure}

Electrical isolation of the ADC units is a critical aspect of the independent dual-system design of the Holometer. Each ADC is housed in a separate NI PXIe-1082 chassis connected via a 30-m fiber optic cable to the PCIe backplane of a dedicated workstation, located in a separate building. The central computer receiving the digitized optical signals is the only link between the four chassis. Direct sample clock sharing would cause correlated phase noise through up-conversion of signals, and also violate the stringent electrical isolation required of the two detection systems. Instead, each ADC unit is individually synchronized to a GPS-provided clocking signal. This signal is generated by an NI PXI-6683H timing card installed in each chassis, with each timing card connected to its own GPS receiver. A schematic is shown in Fig. \ref{fig:fast_adc_schematic}. The interchannel decorrelation due to clocking phase noise, assessed in \S\ref{sec:timing_instability}, is shown to be a small effect ($< 10\%$) and is taken into account in the calibration analysis.

\subsection{Real-Time Data Processing Pipeline}
\label{sec:pipeline}

The 32-core central computer runs a custom C++ program which functions as a high-throughput pipeline capable of processing data streams of up to 1.6~GB/s in real time. The software interfaces with the low-level NI device drivers to provide high-level control of the digitization and synchronization hardware. Before entering the pipeline, the eight input signals to the 14-bit ADCs are anti-aliased via an analog filter and sampled at 100~MHz, then digitally decimated to 50~MHz. In the pipeline, the 50MHz time series are divided into accumulation time intervals of a configurable duration, typically 1-2~s, and processed in parallel. 

Discrete Fourier transforms (DFTs) of a configurable length, $N_{\rm DFT}$, are computed from the time series in batches via the FFTW algorithm. The cross-spectral density (CSD) matrix of all signals is then computed from each DFT batch. The signal data in each DFT are weighted by a Hann window, and successive DFT batches are overlapped by 50\% to recover the information lost by the windowing \citep{Welch1967}. Nominally, $N_{\rm DFT}=2^{17}$ is chosen to achieve a spectral resolution of 570~Hz.

As the data are processed, the CSD matrix computed between the signals of each DFT batch is continually accumulated into a running average over the full accumulation time. The final time-averaged CSD matrix is then written to a 10TB drive. As the averaged CSD matrix is only written once per accumulation time, this technique achieves a data compression factor of $>100$ compared to writing the raw time samples directly to disk. Moreover, because the Holometer is searching for a stationary noise background, as opposed to a time-resolved transient or periodic signal, there is no loss of statistical information traded off for this compression.

Achieving this high level of throughput requires code optimization at the level of the hardware architecture. Profiling during development identified memory bandwidth bottlenecks arising from effectively random memory allocation across the four non-uniform memory access (NUMA) nodes. Hand-tuned optimizations are implemented via the hwloc C++ package \citep{hwloc}, which provides programmatic, location-specifiable memory allocation and thread creation.

A remote python management script interfaces with the low-level C++ process via a TCP connection. The script provides a graphical interface for users to monitor and control the data acquisition. It renders configurable real-time plots of the accumulating correlation matrix, and it also serves as the access point for retrieving previously-recorded data from the server archive. Multiple remote client applications can simultaneously connect and monitor the system and the control system can manipulate the recording state.

\section{Instrument Characterization}
	\label{sec:Characterization}
	
Although measures have been taken to isolate the two interferometers from each other, they still share a common
laboratory space, so low-level enviromental correlations are still expected. Enviromental backgrounds coupling to both
instruments can introduce correlated fluctuations in the optical readouts inferred to be length fluctuations. This
section constructs a noise model used to express limits on common radio frequency (RF) backgrounds leaking into
electronics and into the lasers themselves. From this model, the relative sensitivity of correlated DARM noise to the
presence of instrumental, quantum and correlated environmental noise is developed. In the following sections, nomenclature for the interferometer noise model is first introduced. Then, within this framework, auxiliary measurements of environmental noise are used to construct a budget of the spurious contributions to the interferometer correlation measurement.

\subsection{Nomenclature}
\label{sec:nonemclature}
The number of mechanisms that contribute spurious correlations is limited, whereas the number of sources for the noise, or ``noise modes'' coupling through a particular mechanism, is arbitrary and is generally unknown. To represent this, the statistical time series measurements are decomposed into independent noise modes, appearing as basis vectors of a linear space. This provides an algebraic description of interferometer output signal time series in terms of CSD measurements.

In this algebra, vectors of white noise modes of unit variance are written as $\Wv[proc.]\tb{inst.}(t)$, where the superscript specifies the coupling process or physical mechanism of the mode and the subscript specifies the instrument, such as which interferometer readout channel. The hat is used to indicate that the variable represents a random signal. While random signals may be indexed by time in derivations, statistical operations such as the power spectrum, $\PSD{\cdot}$, are necessary to relate signals to physical observations. These spectrum operations represent the expectation of finite-duration measurements. While taking time series as input, they may be indexed by either time or frequency.

Interferometer signal outputs are represented as linear combinations of white noise modes acted on by transfer function coefficients, $H\tu{proc.}\tb{inst.}(t, f)$, providing the conversion to physical units. The transfer functions represent linear dynamics of optical, electrical, or mechanical systems that are frequency dependent and possibly changing with time. The time dependence $t$ may be dropped to imply stationary response or static transfer gain. Analogously to the vectors of white noise modes, vectors of noise mode couplings are represented by bolded variables. For brevity, the dot product will be used to denote a pairwise action of transfer functions on noise modes, with the frequency-dependent action of convolution implied by the coefficient multiplication notation. In total, the notation here attempts to concisely navigate between three vector spaces: statistical, noise-modes, and time/frequency. The inner products for the first two arise from the cross-spectral density operation and the dot-product respectively. The time-frequency space will be indexed in this treatment, although it is important in integrating the data and error analysis across frequency/time for scientific test statistics.

The action of transfer functions on white noise modes to produce interferometer signal outputs is illustrated in the below example. The number of internal modes or functions of bolded vector-valued variables is left unspecified to imply the arbitrary (unknown) number of physical sources. However, vector-valued variables must always appear dotted in physical observables, as shown below and in Eq.~\ref{eq:noise_dotted}:
\begin{align}
  \sop{M}\tu{ex}\ifoL(t) &= \vect{H}\tb{1}\tu{A}(f,t) \cdot \Wv[A](t)
                           +\vect{H}\tu{B}\tb{1}(f,t) \cdot \Wv[B]\tb{1}(t)
                           \label{eq:M_ex_1}
  \\   
  \sop{M}\tu{ex}\ifoT(t) &= \vect{H}\tb{2}\tu{A}(f,t) \cdot \Wv[A](t)
                             +\vect{H}\tu{B}\tb{2}(f,t) \cdot \Wv[B]\tb{2}(t)
                           \label{eq:M_ex_2}
\end{align}
In the above equations, $\sop{M}\tu{ex}\ifoL(t)$ and $\sop{M}\tu{ex}\ifoT(t)$ are statistical variables representing the output signal time series. They are single (non-array) values despite being constructed from many internal modes through the dot products. These signals express random variables at each moment in time, where their values may have correlations across time and contain non-white spectral content. Additional letters signify different types of signals and noise, as summarized in Table~\ref{tab:notation}.

Cross-spectra measure the correlations of the data channel time series in the frequency domain. White noise modes of the same annotation are perfectly correlated, while those of different annotations are uncorrelated. This is expressed through the following identities:
\begin{align}
    \CSD{\Wv[A]\tb{1}, \Wv[A]\tb{1}} &= \mathbf{1}
    &\CSD{\Wv[A]\tb{1}, \Wv[A]\tb{2}} &= \mathbf{0}
    \nonumber
    \\ \CSD{\Wv[A]\tb{1}, \Wv[B]\tb{1}} &= \mathbf{0}
    & \CSD{\Wv[C], \Wv[C]} &= \mathbf{1}
    \label{eq:noise_mode_algebra}
\end{align}
Blank subscripts indicate common background modes that correlate between the two instruments. The linearity of the CSD statistic allows Eqs.~\ref{eq:noise_mode_algebra} to reduce CSD expressions to inner products of physical transfer functions. In the case of Eqs. \ref{eq:M_ex_1} and \ref{eq:M_ex_2}, the CSD of the detector signals becomes
\begin{align}
  \CSD{\sop{M}\tu{ex}\ifoL, \sop{M}\tu{ex}\ifoT; f, t} 
	&= \vect{H}\tb{1}\tu{A}(f, t) \cdot \conj{\vect{H}\tb{2}\tu{A}(f, t)}
    \label{eq:noise_dotted}
\end{align}
Since the coupling coefficients of the second signal argument are conjugated, the CSD operation is a sesquilinear form in the basis of noise modes. These properties will be used to construct a noise budget of spurious correlations in the following sections.

Experimentally, the data system implements a statistical estimator of the CSD matrix through windowed Fourier transforms and Welch-method accumulation (see \S\ref{sec:pipeline}). The finite sampling rate and measurement intervals are expressed through the additional time and frequency indices in $\CSD{\cdot, \cdot; t, f}$ and $\PSD{\cdot; t, f}$ operations. The indices appear continuous for brevity rather than indexing sample ranges or discrete frequency bins. The time index is necessary to express the changing internal state of the interferometer and corresponding calibration effects (see \S\ref{sec:calibration}), but will be dropped when the spectrum may be idealized as stationary. For succinctness, the frequency index will also be dropped in the following sections, where frequency dependence is implied.

\begin{table}[!tp]
	{\renewcommand{\arraystretch}{1.3}
	\centering
    \begin{tabular}{| l | p{5cm} |}
    \hline
      {\bf Symbol}
      & {\bf General Definition} \\ \hline
      \raisebox{-.1em}{$\sop{M}\tu{proc.}\tb{inst.}(t)$}
      & Calibrated signal time series (units of differential length). \\ \hline
      \raisebox{-.1em}{$\sop{V}\tu{proc.}\tb{inst.}(t)$}
      & Uncalibrated signal time series (units of sensor voltage). \\ \hline
      \raisebox{-.3em}{$\vsop{W}\tu{proc.}\tb{inst.}(t)$}
      & Vector of white noise modes of unit variance. \\ \hline
      \raisebox{-.3em}{\parbox[t]{1.5cm}{
        $\holoMag(f)$ \\[.2em]
        $N\tu{proc.}\tb{inst.}(f, t)$ \\[.2em]
        $\vect{\Gamma}\tb{inst.}(f, t)$ \\[.2em]
        $\vect{\Lambda}\tb{inst.}(f, t)$ \\[.2em]
        $\vect{H}\tu{proc.}\tb{inst.}(f, t)$
      }}
      & Vector of transfer functions coupling noise modes to the system. Each symbol corresponds to a specific coupling mechanism defined in the text in units of differential length or sensor voltage. \\ \hline
    \end{tabular}
\caption{Summary of the naming conventions used by \S\ref{sec:Characterization} - \ref{sec:performance}.}
\label{tab:notation}}
\end{table}

\subsection{Interferometer Noise Model}
    \label{sec:noise_model}
    
The output beam of each interferometer is split between two photodetectors whose readout signals are calibrated to units of differential length (see \S\ref{sec:calibration}). Using the notation introduced in the previous section, the interferometers themselves will be indexed by subscripts 1 and 2, while their photodetectors will be indexed by A and B.  The signal measured by each detector is a linear superposition of noise modes,
\begin{align}
  \sop{M}\ifoL[A]\michdl = \bigg(\hspace{1em} &\hspace{1.2em}\holoMag \, \Wholo  
  && \text{[Exotic noise]} 
  \\ +\, & N\tu{shot}\ifoL[A] \, \W\tu{shot}\ifoL[A]  \nonumber
  && \text{[Shot Noise]}
  \\ +\, & \VNL\,\Vdot\,\WBGL  \nonumber
  && \text{[System Noise]}
  \\ +\, & \VBGL\,\Vdot\,\WBG \hspace{1em}\bigg)
  && \text{[Environmental noise]}\nonumber
\end{align}
The ``exotic noise'' mode is common to both interferometers, while the ``shot noise'' modes are independent for each photodetector. The ``system noise'' modes include amplitude and phase modulations imprinted on the light from sources unique to one interferometer. Modulations common to both interferometers are termed ``environmental noise.'' All of the noise modes are white in frequency and unitless. The transfer functions, shown above, convert these modes into differential length fluctuations measured at the antisymmetric port of the interferometer.

The two photodetector signals of each interferometer are recombined with an optimal weighting to minimize the noise density in the combined channel. For concreteness, an idealized case in which both detectors have equal sensitivity 
is shown below.
\begin{align}
    \sop{M}\michdl\ifoL &= \frac{1}{2}\left( \sop{M}\michdl\ifoL[A] + \sop{M}\michdl\ifoL[B] \right)
\end{align}
With the frequency dependence implicit for brevity, the CSD of the sum channels of the two interferometers is, 
\begin{align}
    \CSD{\sop{M}\michdl\ifoL, \sop{M}\michdl\ifoT} &= \abs{\holoMag}^2 + \VBGL\Vdot\conj{\VBGT} \;,
  \label{eq:two-ifo-cross-budget}
\end{align}
measures the exotic correlated noise power, plus any environmental noise leakage, which can introduce bias. The independent shot noise and system noise in the detector channels average away in the limit of long integration times, leaving only the correlated noise component. Analogously, the CSD of the two individual detector channels of one interferometer,
\begin{align}
    \CSD{\sop{M}\michdl\ifoL[A], \sop{M}\michdl\ifoL[B]} = \abs{\holoMag}^2 + |\VBGL|^2 + |\VNL|^2 \;,
  \label{eq:single-ifo-cross-budget}
\end{align}
is sensitive to the terms of Eq.~\ref{eq:two-ifo-cross-budget} as well as to the system noise of the interferometer, which is common to both detectors.

Each system noise and environmental noise term, represented as a vector of couplings to a pool of source noise modes, can be further decomposed into the physical mechanisms through which noise couples into optical signals at the detectors. The environmental noise common to all detectors on both interferometers is represented as the linear superposition
\begin{align}
\label{eq:env_decomp}
     \VBGL &= \VBGL[AM] + \VBGL[PM] + \VBGL[RF] + \VBGL[X] \;.
\end{align}
Similarly, the system noise common to all detectors on a single interferometer is represented as
\begin{align}
\label{eq:sys_decomp}
     \VNL &=  \VNL[AM]  + \VNL[PM]  + \VNL[RF] + \VNL[KT] + \VNL[X] \;.
\end{align}
In Eqs.~\ref{eq:env_decomp} and \ref{eq:sys_decomp}, the superscripts denote the physical mechanism for each coupling, as listed below.
\begin{description}
    \item[AM] Amplitude noise in the laser
    \item[PM] Phase/frequency noise in the laser
    \item[RF] Radio pickup in the readout electronics and cables
    \item[KT] Thermal noise in the mirror substrates
    \item[X] Any unknown coupling mechanism
\end{description}
While each term represents a different mechanism of coupling noise into the instruments, they are all expressed in terms of their apparent DARM effect at the antisymmetric port.

For all cross-spectra, the statistical uncertainty is determined from the two auto-spectra together with the integration time, $T\tb{span}$, and bandwidth, $F\tb{span}$, of the measurement. Accounting for the small loss of degrees of freedom from the use of the Welch method (see \S\ref{sec:pipeline}), the variance on the CSD estimator (either real or imaginary components, independently), degraded by the Welch overlap \citep{Harris1978} is
\begin{align}
     \text{Var}\left[ \CSD{\sop{M}\michdl\ifoL, \sop{M}\michdl\ifoT} \right] &= 
\frac{\text{PSD}[\sop{M}\michdl\ifoL] \, \text{PSD}[\sop{M}\michdl\ifoT]}{0.75 \, T\tb{span} F\tb{span}} \;.
  \label{eq:CSD_statistics}
\end{align}
Because the sensitivity of the interferometers varies over long integration time due to alignment drift, the power spectra in Eq.~\ref{eq:CSD_statistics} are not constant for each independent sample in the data set. However, they are stable for intervals sufficiently long to determine the local statistical weight. The data are combined using these weights.

\subsection{Interferometer Cross-Spectrum}
	\label{sec:data-integration}

The experimental correlation measurement of differential length fluctuations, expressed by
Eq.~\ref{eq:two-ifo-cross-budget}, is shown in Fig.~\ref{fig:AS_results} for the first 145~hours of first-generation
Holometer data. The top panel shows the magnitude of the interferometer cross-spectrum at 3.8~kHz resolution, while the
lower two panels show the real part used to test for quantum-geometrical noise at 3.8~kHz and then 64~kHz resolution. The $1\sigma$ statistical uncertainty is denoted by the black curve in the top panel and by the black error bars in the bottommost panel. The error bars are suppressed in the center plot for clarity.

The figure indicates the presence of some correlated noise below 1~MHz, requiring analysis of environmental background contributions and consequent selection of time- and frequency-domain vetoes (see \S\ref{sec:quality_control}). After background analysis, each non-vetoed real data point in Fig.~\ref{fig:AS_results} is consistent with zero correlation. However, this presentation does not test for any particular quantum-geometrical noise model, as the frequency bin integrations do not weight by any model noise spectrum. Scientific analysis of this data set is separately performed in \cite{HoloPRL} to place constraints on a candidate model. Final analysis of the full 704-hour data set, which has the statistical power to constrain a broader class of models, is forthcoming.

\begin{figure}[!tp]
  \centering
  \includegraphics[width=1\linewidth]{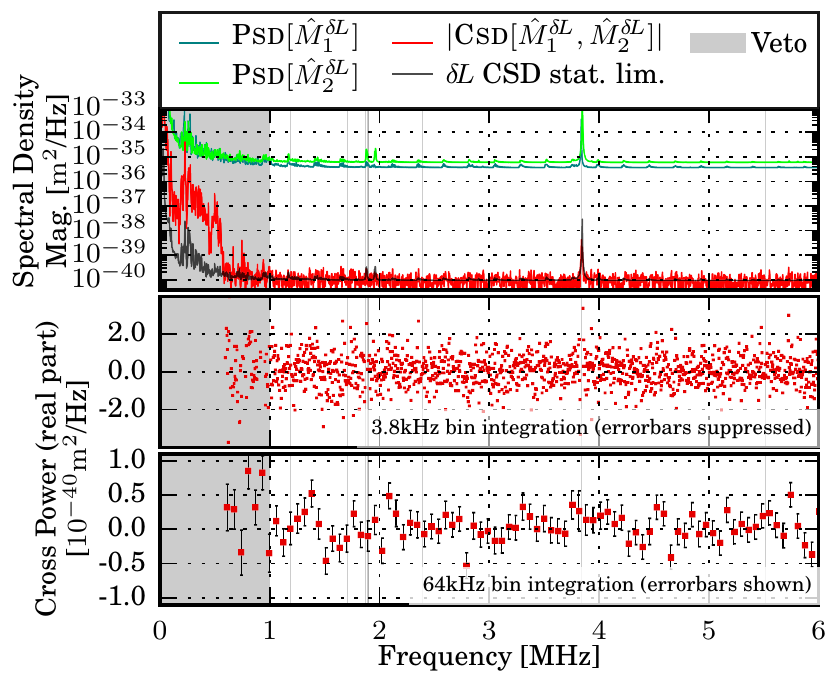}
  \caption{
    Measured power and cross spectra of the first-generation interferometer DARM channels after 145~hours of integration. In the top panel, the two green curves denote each interferometer's respective power spectrum. The red curve denotes the magnitude of the cross-spectrum of the two interferometer output signals, $\abs{\holoMag}^2 + \VBGL\Vdot\conj{\VBGT}$. The $1\sigma$ level of statistical uncertainty is denoted by the black curve. In the center panel, the real part of the cross-spectrum used to test for quantum-geometrical noise is shown at 3.8~kHz resolution, with the statistical error bars suppressed for clarity. In the bottom panel, 64-kHz bin integration is used to show 4x lower noise, with each non-vetoed point statistically consistent with zero.
  }
  \label{fig:AS_results}
\end{figure}

\subsection{Environmental Background}
    \label{sec:env_background}

As shown in Eq.~\ref{eq:two-ifo-cross-budget}, the presence of spurious environmental correlations biases the interferometer cross-spectrum as an estimator of the quantum-geometrical noise. To bound the magnitude of the background leakage term, $\VBGL\Vdot\conj{\VBGT}$, the noise entering through each known physical coupling mechanism is monitored during data collection and cross-correlated with the high-speed photodetector outputs. {\it Ex situ} measurements are used to calibrate the external pickoff/testpoint monitors to units of differential length at the antisymmetric port. The linear combination of background measurements is used to construct the upper limit of correlated background contributions,
\begin{align}
      \label{eq:budget-measurements}
  (\VBGL - \VBGL[X])\Vdot\conj{\VBGT} = 
    \bigg(\hspace{1em} & 
     \CSD{\sop{M}\tu{AM}\ifoL, \sop{M}\michdl\ifoT} 
  \\+ \;&\CSD{\sop{M}\tu{PM}\ifoL, \sop{M}\michdl\ifoT} \nonumber
  \\+ \;&\CSD{\sop{M}\tu{RF}\ifoL, \sop{M}\michdl\ifoT} \nonumber
      \hspace{1em} \bigg)\;.
\end{align}
In the above equation, the CSDs respectively represent the components of amplitude noise, phase noise, and dark noise in interferometer 1 that correlate with the output signal of interferometer 2. Only the coupling of the unknown backgrounds to each other, $\VBGL[X]\Vdot\conj{\VBGT[X]}$, cannot be determined by this technique. Accounting for the known coupling mechanisms, as described below, the noise budget, excluding the $\VBGL[X]\Vdot\conj{\VBGT[X]}$ term is small ($<5\%$) at frequencies below 8~MHz except in narrow bands that are vetoed using the chi-square statistic showing excess correlation with auxiliary channels.

The spurious correlation budget in Eq.~\ref{eq:budget-measurements} requires a series of measurements between antisymmetric-port and auxiliary sensors, with the auxiliary data calibrated to the effect on the differential length signal at the antisymmetric port. The following sections detail the experimental setups used to make and calibrate the {\it in situ} measurements for the environmental background budget, as well as the noise model employed for each sensor. All of the sensors are shot noise-limited and the spurious correlation limits are dominated by the statistical uncertainties of the budget estimates rather than by detected backgrounds.

\subsubsection{Laser Amplitude Noise Limits}
	\label{sec:am-mitigation}

To measure the correlated amplitude noise background of the lasers, $0.5\%$ of the beam power is picked off after all of the active electro-mechanical elements between the laser and interferometer and directed onto a New Focus 1811 photodetector inside the Mach-Zehder (with blocked arm) of Fig. \ref{fig:L_laser_launch}. The small pickoff power gives this detector poorer shot noise-limited sensitivity to amplitude modulations of the laser. However, the amplitude modulations leaking through to the interferometer output are attenuated by the storage cavity. These effects balance to give the pickoff monitor an order of magnitude better sensitivity to amplitude modulations than the antisymmetric-port detectors, allowing it to budget a spurious systematics limit with 1\% of the standard error of the principal antisymmetric-port cross-correlation when monitored for the same duration.

The amplitude noise monitor measures the linear combination of laser amplitude noise, environmental noise, and shot noise,
\begin{align}
\sop{M}\tu{AM}\ifoL = 
   \VNL[AM] \Vdot \WBGL + \VBGL[AM] \Vdot \WBG 
  + N\tu{AM}\ifoL \W\tu{shot}\ifoL[,AM] \;.
\end{align}
This signal is calibrated to its effect on the differential length signal at the antisymmetric port as
\begin{align}
  \sop{M}\tu{AM}\ifoL = H\michdl\tb{1,AM} \sop{V}\tu{AM}\ifoL \;,
\end{align}
where $\sop{V}\tu{AM}\ifoL$ is the raw detector readout signal and $H\michdl\tb{1,AM}$ is the optical transfer function of amplitude sidebands to power modulations at the output port. The antisymmetric port-referred signal calibration is measured {\it ex situ} using the second pickoff photodetector to perform a lock-in measurement,
\begin{align}
H\michdl\tb{1A,AM}
  = \frac
    {\CSD{\sop{M}\ifoL\michdl, \sop{V}\tu{AM}\ifoL[B]}}
    {\CSD{\sop{V}\tu{AM}\ifoL[A], \sop{V}\tu{AM}\ifoL[B]}} \;.
\end{align}
The passive amplitude noise of the laser is sufficient over a one-hour integration of these CSD statistics to provide a measurement of the transfer function during typical operation of the interferometer. This transfer function agrees with optical models of the pickoff and cavity pole, with additional frequency shape from higher-order modes leaking at frequencies set by Eq.~\ref{eq:hom_freqs}.

During commissioning, measured antenna correlations with these amplitude noise monitors are used to validate procedures to isolate the laser and its pump diode from the RF environment. These isolation measures include the following. The power cable of each laser was wound through lossy Ferroxcube TX74/39/13-3C81 ferrite toriods, which function as common-mode electrical chokes. An isolating transformer was installed between the power supply of each laser and the Fermilab power mains. A direct safety shutoff connector on each laser, found to be introducing common electrical noise, is replaced with an AC-power kill switch safety mechanism to avoid any connections to those terminals. Likewise, the power cables of the external electrical components on the laser table were individually wound through common-mode chokes and connected through isolating transformers. This environmental isolation process has been iterated until antenna correlation with the laser amplitude noise was acceptably small.

The measured limits on the correlated amplitude noise of the lasers are shown in Fig.~\ref{fig:ASLIM_IMON_lin}. The blue and violet curves are the amplitude noise background terms, $\VBGL[AM]\Vdot\overline{\VBGT}$ and $\VBGT[AM]\Vdot\overline{\VBGL}$, respectively. The shaping of the noise limits reflects the transfer function of the Fabry-Perot cavity, which has peak transmission at the free spectral range, $c/2L$, and its harmonics. The excess cross-spectral power is larger at those frequencies because the laser amplitude noise is less attenuated. The red curve is the measured cross-spectrum of the interferometer output signals, with the black curve indicating the statistical expectation for uncorrelated noise. Below 600~kHz, the divergence of these curves indicates significant excess correlated power. The figure shows that laser amplitude noise, represented by the bottom two curves, is not the dominant source of this excess. If it were, the lowest curves would overlay the red curve. Nevertheless, the chi-square statistic in the lower panel of the figure shows a significant detection of correlated amplitude noise at frequencies below the 1MHz signal band.

\begin{figure}[!tp]
  \centering
  \includegraphics[width=1\linewidth]{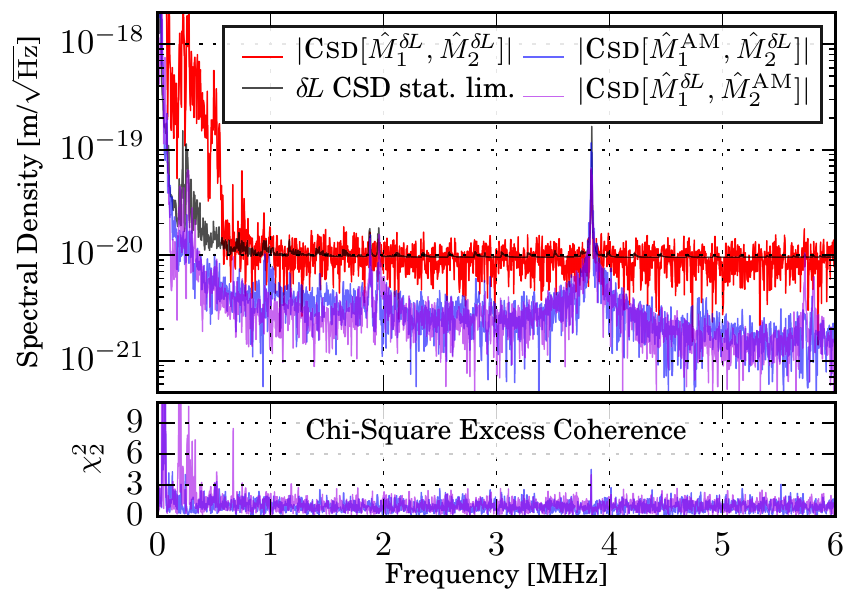}
  \caption{Measured limits on the correlated amplitude noise background of the lasers. The blue and violet curves are the amplitude-noise background terms, $\VBGL[AM]\Vdot\overline{\VBGT}$ and $\VBGT[AM]\Vdot\overline{\VBGL}$, respectively. The red curve is the cross-spectrum of the two interferometer output signals, $\abs{\holoMag}^2 + \VBGL\Vdot\conj{\VBGT}$ (at the plotted sensitivity level, the exotic $\abs{\holoMag}^2$ term is negligible). The reduced $\chi^2$ statistic indicates the level of statistical certainty of noise detection in the background limit. Values $>3$ are indicative of significance at the 95\% confidence level.}
  \label{fig:ASLIM_IMON_lin}
\end{figure}

\subsubsection{Laser Phase Noise Limits}
	\label{sec:pm-mitigation}

At frequencies above the 350-Hz cavity pole, where the response is flat in phase, the PDH error signal (see \S\ref{sec:carm_loop}) measures the correlated phase noise background of the lasers. The PDH error signal is shot noise-limited to 3~MHz, but it is not optimized for phase sensitivity. Dynamic range limits of the {\it in situ} hardware prevented strong modulation depths or increased optical power on the detector, and, to prevent introducing additional RF contamination into the laser frequency actuator, additional electronics could not be added. However, due to the small arm length imbalance (Schnupp asymmetry \citep{schnupp}), the conversion of phase noise to amplitude noise in the Michelson interferometers is weak. These effects balance to give the PDH signals nearly an order of magnitude better sensitivity to phase modulations than the antisymmetric-port detectors, allowing it to budget a spurious systematics limit with 1\% of the standard error of the principal antisymmetric-port cross-correlation.

The PDH error signal measures the linear combination of phase noise, environmental noise, and shot noise,
\begin{align}
\sop{M}\tu{PM}\ifoL = 
   \VNL[PM] \Vdot \WBGL + \VBGL[PM] \Vdot \WBG 
  + N\tu{PM}\ifoL \W\tu{shot}\ifoL[,PM] \;.
\end{align}
This signal is calibrated to its effect on the differential length signal at the antisymmetric port as
\begin{align}
  \sop{M}\tu{PM}\ifoL = H\michdl\tb{1,PM} \sop{V}\tu{PM}\ifoL \;,
\end{align}
where $H\michdl\tb{1,PM}$ is the optical transfer function of phase modulations to power modulations at the output port. The antisymmetric port-referred PDH signal calibration is measured using an electro-optic modulator (EOM) to directly drive the laser phase. The drive signal, $\sop{V}\tu{inj}\ifoL$, is swept in frequency and the calibration measured through the lock-in estimator
\begin{align}
H\michdl\tb{1,PM}
  &= \frac
    {\CSD{\sop{M}\ifoL\michdl, \sop{V}\tu{inj}\ifoL}}
    {\CSD{\sop{V}\tu{PM}\ifoL, \sop{V}\tu{inj}\ifoL}} \;.
\end{align}
This calibration matches interferometer optical models of the PDH signal. A Mach-Zehnder interferometer (see Fig.  \ref{fig:L_laser_launch}) with a 2-foot arm length imbalance simultaneously detects the phase and amplitude modulations driven by the EOM. Although it is less sensitive to phase modulations, its output signals are cross-correlated with the PDH signal to bound the residual amplitude modulation from the EOM drive.

During commissioning, mitigation of laser phase noise correlations have proceeded similarly to the amplitude noise mitigation (see \S\ref{sec:am-mitigation}). The PDH error signals are correlated with RF antennas to identify locations in the signal chain susceptible to environmental leakage. The active control-loop electronics were found to drive phase noise in the laser, and the DC/RF splitting of the PDH signal has been found to introduce ambient RF pickup in the channel lines. To suppress the electronics noise, common-mode electrical chokes were installed before and after each cable connection between electrical components. Moreover, to minimize ambient pickup, Heliax isolating corrugated copper cables are used to transport the environmental monitoring signals from the laser tables to the data acquisition system. This process was iterated until antenna correlation with the laser phase noise is acceptably small.

The measured limits on the correlated phase noise of the lasers are shown in Fig.~\ref{fig:ASLIM_PDH_lin}. The blue and violet curves are the phase noise background terms, $\VBGL[PM]\Vdot\overline{\VBGT}$ and $\VBGT[PM]\Vdot\overline{\VBGL}$, respectively. The red curve is the measured cross-spectrum of the interferometer output signals, with the black curve indicating the statistical expectation for uncorrelated noise. Unlike the case of amplitude noise, where only a few narrow frequencies exhibit significant correlation between interferometers, laser phase noise correlation is significant for a large portion of the spectrum below 600~kHz. It is clear from the figure that laser phase noise is the dominant source of low-frequency correlation, as the phase noise limits nearly saturate the correlation budget shown by the red curve.

\begin{figure}[!tp]
  \centering
  \includegraphics[width=\linewidth]{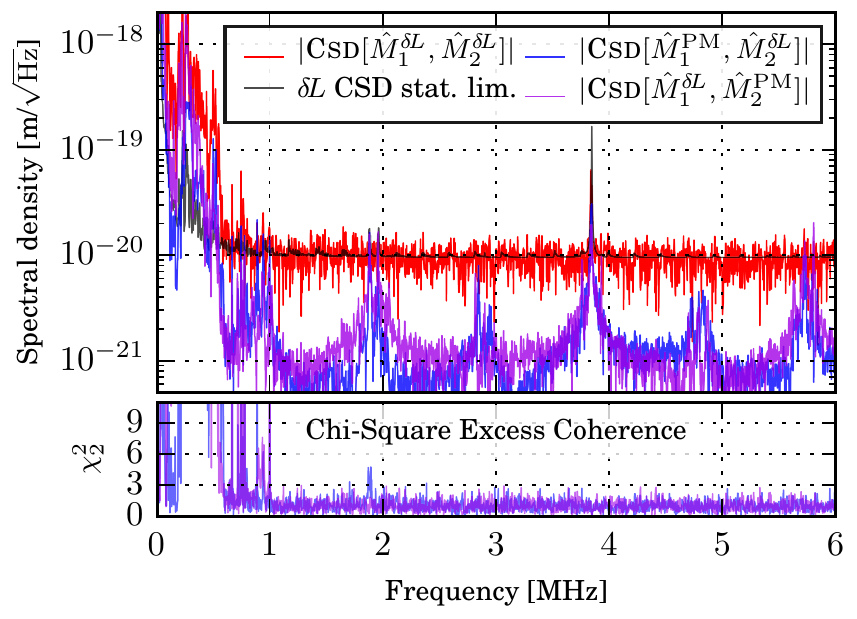}
  \caption{Measured limits on the correlated phase noise background of the lasers. The blue and violet curves are the phase-noise background terms, $\VBGL[PM]\Vdot\overline{\VBGT}$ and $\VBGT[PM]\Vdot\overline{\VBGL}$, respectively. The red curve is the cross-spectrum of the two interferometer output signals, $\abs{\holoMag}^2 + \VBGL\Vdot\conj{\VBGT}$ (at the plotted sensitivity level, the exotic $\abs{\holoMag}^2$ term is negligible). The reduced $\chi^2$ statistic indicates the level of statistical certainty of noise detection in the background limit. Values $>3$ are indicative of significance at the 95\% confidence level.} 
  \label{fig:ASLIM_PDH_lin}
\end{figure}

\subsubsection{Readout Electronics Noise Limits}

The signal-readout chains, where RF pickup can occur in front-stage amplifiers, are a major entry point for environmental contamination. They are also the simplest to measure against RF antennas, as measurements can be made passively without an operational interferometer. Without the additional photon noise in the detectors, long integrations of the ``dark noise,'' calibrated to units of differential length, can probe far below the expected sensitivity of the instrument. The signal-chain electronics are isolated using common-mode electrical chokes installed before and after all signal lines to the readout ADCs, and all of the power supplies are choked and connected through isolating transformers. Dark noise correlation measurements can not be made {\it in situ}, but they have been regularly repeated during data collection to ensure that the {\it ex situ} limits are representative of the long-term operating environment. Multi-hour CSD measurements between the dark readout channels firmly establish that the dark noise couplings, $\VBGL[RF]$ and $\VBGT[RF]$, make negligible contributions to the correlated interferometer noise budget ($<<1\%$).

\subsection{Thermal Noise in End Mirrors and Beamsplitter}
	\label{sec:thermal_noise}

The cross-spectrum of the two detectors of one interferometer (Eq.~\ref{eq:single-ifo-cross-budget}) is sensitive to the system noise component. Above 1~MHz, the measured noise is dominated by thermally-excited vibrations of the end mirrors and beamsplitter, represented by the coupling $\VNL[KT]\Vdot\WBGL$. This additive thermal noise is of small consequence for the cross-interferometer measurement, as its magnitude is comparable to that of the shot noise and it does not correlate between the two optical systems. However, the detection of the thermal mirror noise does demonstrate the ability of the Holometer to detect differential motion, below shot noise, at radio frequencies. The thermal noise is used to validate the calibration of each interferometer to within a factor of two, limited by the confusion of thermally-driven angular and longitudinal noise in the antisymmetric-port response. 

\subsubsection{Spectral Model of Thermal Excitations}

Fig.~\ref{fig:csd_same_ifo} shows the magnitude of the single-interferometer CSD of one interferometer. The noise spectrum is punctuated with a regular set of spectral lines spaced approximately every $226$~kHz. The frequencies of the peak-amplitude lines are consistent with harmonics of the round-trip sound-crossing time in the end mirrors and beamsplitter (mechanical etalons). Each quartz optic, of thickness $d=12.7$~mm, has a sound speed of $v_s=5720$ m/s. As planar vibrational modes in a cylinder have a fundamental frequency of $v_s/2d=225$~kHz, these lines are identified as thermal excitations of the fundamental planar mode and its higher harmonics. Since the smallest dimension of the optics is their thickness, the other vibrational modes with radial and azimuthal dependence will have different, more closely-spaced excitation frequencies. These modes are interpreted to correspond to the remaining system of peaks seen regularly spaced in the figure.

\begin{figure}[!tp]
    \centering
    \includegraphics[width=1\linewidth]{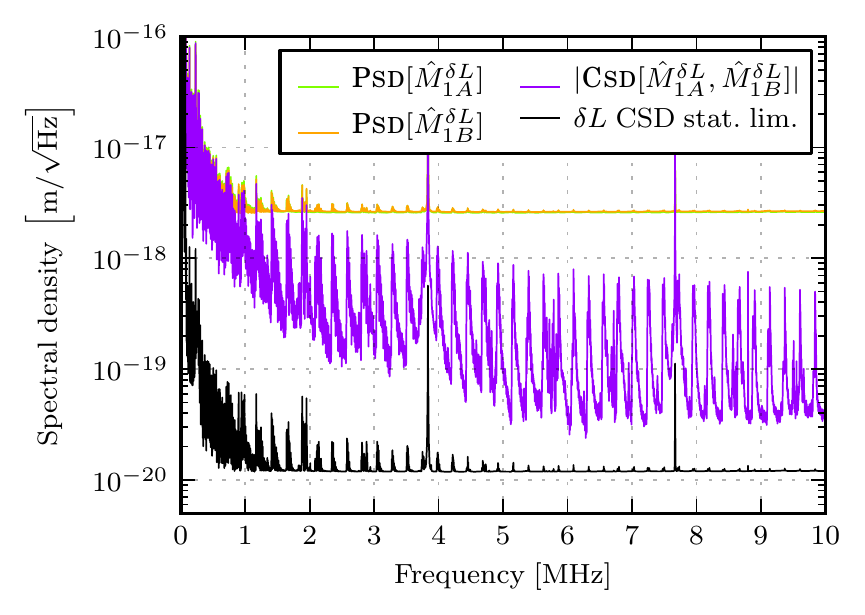}
    \caption[Thermal noise of one interferometer measured via the single-interferometer CSD.]{Thermal noise of one interferometer measured via the single-interferometer CSD. The noise spectrum is punctuated with a regular set of spectral lines spaced approximately every $226$~kHz, consistent with harmonics of the round-trip sound-crossing time in the end mirrors and beamsplitter. The peak-amplitude lines are identified as thermal excitations of the fundamental planar vibrational mode and its higher harmonics. The remaining system of regularly-spaced peaks is identified as other vibrational modes with radial and azimuthal dependence.}
    \label{fig:csd_same_ifo}
\end{figure}

In thermal equilibrium, each vibrational mode of the optical elements has energy $k_bT/2$ at temperature $T$. Each mode is an acoustic standing wave with total energy $\frac{1}{2}m \omega^2 \langle x^2\rangle$ with $\langle x^2\rangle$ the RMS amplitude. At the surface of the mirror, the mode thus drives a mean-square surface motion of
\begin{align}
  \langle x^2\rangle = \frac{k_bT}{m\omega^2} \;,
	\label{eq:thermal_rms}
\end{align}
where $m=56$~g is the mass of the mirror and $\omega$ is the angular frequency of the normal mode oscillation. The power spectral density of each resonant response is a Lorentzian curve,
\begin{align}
L(f) = h \; \frac{(\Gamma/2)^2}{(f-f_0)^2+(\Gamma/2)^2} \;,
\end{align}
where $h$ is the height of the peak, $f_0$ is the central frequency, and $\Gamma$ is the width. When the data acquisition system is operated at its highest resolution, 23 Hz, each line is individually resolved (the nominal operating resolution is 570~Hz; see \S\ref{sec:pipeline}). The integral of an individual Lorentzian,
\begin{align}
\langle x^2\rangle = \int L(f) \, df = \frac{h\Gamma}{2\pi} \;,
\end{align}
is equal to the mean-square longitudinal displacement.

In principle, these vibrational modes provide a set of direct, naturally-occurring calibration lines spanning the entire measurement band. Their application is more limited in practice, however, by the confusion of thermally-driven angular and longitudinal noise in the antisymmtric-port response. Even so, the thermal excitations do provide an approximate validation of interferometer calibrations obtained by indirect methods, as discussed below.

\subsubsection{Calibration Using Thermal Noise}

Two methods are used to check indirectly-determined interferometer calibrations (see \S\ref{sec:calibration}) against the expected mean-square thermal motion. The first method directly compares the integrated power in a single calibrated line to that which is expected under the thermal model at equilibrium, as given by Eq.~\ref{eq:thermal_rms}. However, because vibrational modes with azimuthal and radial dependence do not align with 100\% efficiency to the optical path of the interferometer arm, their plane waves are expected to contribute less than $k_bT/m\omega^2$ to the longitudinal length fluctuation spectrum. In practice, it is difficult to identify purely longitudinal modes due to the complicating effects of the mounting and support structure. While many lines are measured to contribute less than $k_bT/m\omega^2$, no line has ever been found to {\it exceed} the thermal $k_bT/m\omega^2$ bound, which would be indicative of an overestimated indirect calibration sensitivity. Because some lines do nearly saturate the $k_bT/m\omega^2$ bound, the thermal measurements made via this method are concluded to be consistent with the indirect interferometer calibrations.

The second method integrates each single-interferometer cross-spectrum over a band of many etalons, subtracting the residual non-thermal background. Unlike the first method, computing the band-integrated noise power does not require the thermal lines to be resolved. The integrated thermal noise power is a proxy for time stability of the calibration. Although the intrinsic thermal noise power is unknown, its inferred value should remain constant in time, at constant temperature (which is maintained; see \ref{sec:end_mirrors}), if the interferometer calibration is stable. However, the etalon correlations in practice measure an unknown combination of longitudinal and angular noise power, and the angular noise component is dependent on alignment. As such, time drift in the calibration accuracy cannot be reliably decoupled from angular alignment drift. Over the entire 704-hour span of the data, this statistic varies by a factor of approximately two, placing a limit on the maximum possible drift in calibration accuracy. The methods of \S\ref{sec:performance} are indirect but place much stronger limits.

\section{Instrument Calibration}
\label{sec:calibration}

The high-speed data processing pipeline outputs a series of 1-second averaged cross-spectrum matrices between the
antisymmetric-port photodetector readouts (see \S\ref{sec:pipeline}) as well as auxiliary channels. These spectra are
generated from the time series of photodiode electronics voltages, which must be calibrated to physical units of
differential length. The conversion accounts for overall gains including the optical length sensitivity and the quantum
efficiency, as well as frequency-dependent gains such as the photodiode circuit transimpedance. The following sections
describe the calibration procedure. Following the notational convention introduced in \S\ref{sec:nonemclature},
convolution of impulse responses in the time domain will be notated as multiplication of transfer functions for brevity.
All system dynamics, in practice, are measured as frequency-domain transfer functions using ratios of finite
cross-spectral density measurements.

\subsection{Overview}

While the experimental noise is expected to be stationary, in the voltage measurement basis the calibration gains slowly drift over time. This drift arises from varying internal state of the interferometers, where the dominant source is uncontrolled alignment degrees of freedom. The voltage signal of each high-frequency (HF) readout channel can be represented as a stationary DARM noise projected through the time-varying transfer function of the instrument,
\begin{align}
\sop{V}\ifoL[A,HF]\michas &= H\ifoL[A,HF]\michdl(f,t) \; \sop{M}\ifoL[A]\michdl
\end{align}
The transfer function $H\ifoL[A,HF]\michdl(f,t)$, mapping differential length changes in the interferometer into the measured voltage response, will be referred to as the calibration of channel 1A. In this language, each cross-spectrum measurement
\begin{align}
&\CSD{\sop{V}_{\rm 1A,HF}\michas, \sop{V}_{\rm 2A,HF}\michas; f,t} \nonumber\\
	&\hspace{2em}= H_{\rm 1A,HF}\michdl(f,t) \;\, \conj{H_{\rm 2A,HF}\michdl(f,t)} \,\; \CSD{\sop{M}_{\rm 1A}\michdl, \sop{M}_{\rm 2A}\michdl; f}
\label{eq:CSD_cal}
\end{align}
must be calibrated by dividing out the transfer functions of the two readout channels.

In principle, the channel calibrations could be directly measured by dithering the PZT-actuated end mirrors with a known injection signal. While standard practice in gravitational wave detectors, direct calibration techniques do not carry over into the Holometer design, because mechanical resonances limit the stable operation bandwidth of the PZT drives to $<1.5$~kHz (see \S\ref{sec:pzt_drive}). The unique, radio frequency requirements of the Holometer led to the development of an indirect calibration ladder to derive the HF response of a Michelson interferometer. This technique, described in the following sections, circumvents the mechanical drive limitations encountered by direct methods.

To simplify notation, the readout channel labels will hereafter be dropped from all terms. The following procedure for calibrating one channel applies analogously to all channels. Furthermore, many of the voltage signals and measurements should have additional electronics response terms (e.g., cabling), which are omitted for brevity. These terms are accounted for in the real calibration measurements, and may be regarded as having been absorbed into the terms appearing below.

\subsection{Indirect Approach}
	\label{sec:indirect_calibration}

The transfer function of any linear system can be decomposed into the product of transfer functions of its individual components. Accordingly, the calibration decouples as
\begin{align}
\label{eq:cal_decomp}
\HcalHF(f,t) &= \HdaqHF(f) \; \Hifo(f,t)
\end{align}
where $\Hifo(f,t)$ (units W/m) is the alignment-dependent, time-varying transfer function of the interferometer and $\HdaqHF(f)$ (units V/W) is the stationary transfer function of the HF signal-readout chain. In \S\ref{sec:pr_as_response} a numerical model of the Fabry-Perot cavity is used to demonstrate that the interferometer transfer function is constant to $2\%$ above $1$~kHz. Invoking this result, Eq.~\ref{eq:cal_decomp} is represented as
\begin{align}
\label{eq:HFcal_decomp_f0}
\HcalHF(f,t)
  &= \HdaqHF(f) \;
    \left( \frac{\Hifo(f,t)}{\Hifo(f_0,t)} \right)
    \; \Hifo(f_0,t) \nonumber \\
  &\approx \HdaqHF(f) \; \Hifo(f_0,t)
\end{align}
for all frequencies $f$ and $f_0 \gtrsim 1$~kHz. Eq.~\ref{eq:HFcal_decomp_f0} expresses the underlying principle of indirect calibration. Suppose the end mirrors are differentially dithered at $f_0\approx1$~kHz, just below the maximum frequency of stable actuation, by a known amount. If the calibration is directly measured at $f_0$, then Eq.~\ref{eq:HFcal_decomp_f0} implies that {\it only} knowledge of the detection chain electronics response is required to infer the calibration for all higher frequencies.

The 1kHz drive is below the high-pass input filter of the HF signal-readout chain (see \S\ref{sec:modified_detectors}). However, the low-frequency (LF) readout of each photodetector, simultaneously sampled at 32~kHz by the control system, is sensitive to the injected signal. The transfer function of the LF system is written as
\begin{align}
\label{eq:LFcal_decomp_f0}
\HcalLF(f_0,t) &= \HdaqLF(f_0) \; \Hifo(f_0,t)
\end{align}
where $\HdaqLF(f_0)$ (units V/W) is the transfer function of the LF photodetector channel and the FPGA. Substituting Eq.~\ref{eq:LFcal_decomp_f0} into Eq.~\ref{eq:HFcal_decomp_f0} yields the expression
\begin{align}
\label{eq:HF_cal_final}
\HcalHF(f,t)
&= \left( \frac{\HdaqHF(f)}{\HdaqLF(f_0)} \right) \; \HcalLF(f_0,t)
\end{align}
formulated purely in terms of continuously-monitored control signals and the measurable, stationary transfer functions of the signal-readout chains. The following sections now detail the practical measurement of each of the above terms. A summary of the main sources of statistical and systematic calibration error is presented in Table~\ref{tab:cal_error_sources}. Each of the error sources is described in the following sections.

\begin{table}[!tp]
	{\renewcommand{\arraystretch}{1.3}
	\centering
    \begin{tabular}{| l | l |}
    \hline
      {\bf Error Source}
      & {\bf Relative Magnitude} \\ \hline
      \raisebox{-.1em}{\parbox[t]{3.75cm}{Statistical error on {\it ex-situ} measurements}} & ${<}1\%$ \\[1.3em] \hline
      \raisebox{-.1em}{\parbox[t]{3.75cm}{PZT calibration drift}} & ${<}2\%$ per interferometer \\[.1em] \hline
      \raisebox{-.1em}{\parbox[t]{3.75cm}{Photodetector dependence on DC optical power}} & 4-7\% per detector \\[1.3em] \hline
      \raisebox{-.1em}{Sampling decoherence} & \parbox[t]{3.75cm}{${<}10\%$ below 8~MHz. (quadratic in frequency)}  \\[1.3em] \hline
    \end{tabular}
\caption{Summary of the main sources of statistical and systematic calibration error. Each of these error sources is discussed in \S\ref{sec:low_freq_calib}-\ref{sec:timing_instability}.}
\label{tab:cal_error_sources}}
\end{table}

\subsection{Direct Low-Frequency Calibration $\,H^{\delta L}_{\rm LF}(f_0,t)$}
	\label{sec:low_freq_calib}

The time-varying dynamics of the interferometers are contained purely in the $\HcalLF(f_0,t)$ terms, which are continuously monitored by the control systems via a direct, low-frequency calibration measurement, as described below.

\subsubsection{Calibration Line Injection}
	\label{sec:calibration_line}

The interferometer control systems are configured to continuously inject a calibration line near 1~kHz via a differential end mirror dither. One interferometer injects a signal at 983~Hz and the other at 984~Hz. For diagnostic purposes, each control system also injects dithers at several lower frequencies spaced between 500-800~Hz. Intermodulation studies using these dithers have placed limits on the longitudinal impurity of the DARM drive, because any residual angular motion could bias the calibration line measurement. QPD measurements of the intermodulation power limit calibration bias from angular drive motion to $<2\%$.

The control systems continuously log the 32kHz-sampled time series of the diagnostic and interferometer control signals, including the digitally-generated drive signal, $\Minj$, and the LF readout of each antisymmetric-port photodetector, $\hVLF$. Since the calibration lines are necessarily measured while the interferometers are locked, the differential-mode control loop acts to suppress the drive injection. A loop gain correction, measured from the coherent time series of the drive and internal test points within the loop, is first applied to remove this effect. The data then measure the out-of-loop transfer function of the true DARM drive amplitude to the LF readout of each antisymmetric-port photodetector,
\begin{align}
 \hVLF &= \HcalLF(f_0,t)\;\Minj
\end{align}
The transfer functions are calculated using complex positive-frequency narrowband filters on the data, dividing by the complex drive signal and then downsampling to 16~Hz. The narrowband filters act as Nyquist filters for the downsampling and the ratio normalizes the drive amplitude and shifts modulation to DC.

\subsubsection{Reference to Optical Wavelength}
	\label{sec:reference_to_wavelength}

In order to inject a calibrated displacement, the drive voltage of each PZT system is itself calibrated to the physical length displacement produced by the actuators,
\begin{align}
\Minj &= H\tu{PZT}(f_0) \; \Vinj
\end{align}
A measurement of this transfer function is made using the laser wavelength as an absolute length reference. The measurement is made in single-pass Michelson configuration, where the PRM is misaligned to prevent multiple traversals of light through the interferometer. In this configuration, the input power is constant, and when locked at mid-fringe, exactly half of the light exits through the antisymmetric port. The half-light condition is established by first sweeping DARM through a full fringe, to determine the minimum and maximum values of the Michelson sinusoidal response, $\text{Max}\left[\hVLF\right]$ and $\text{Min}\left[\hVLF\right]$, respectively. The predicted response to a DARM dither about mid-fringe is
\begin{align}
\label{eq:mich_response}
\hVLF/\sop{M}\tu{inj} &= \frac{2\pi}{\lambda} \, \left(\text{Max}\left[\hVLF\right] - \text{Min}\left[\hVLF\right] \right)
\end{align}
where $\lambda=1064$~nm is the laser wavelength. 

The Michelson interferometer is then aligned so that the contrast defect is at a quadratic minimum and locked to half-power, where any residual variation in contrast defect cannot affect the lock-point offset. A differential swept-sine signal, $\Vinj$, is injected via the PZT drive and measured at two test points simultaneously to infer the loop gain and the optical response at the antisymmetric port. From these, the open-loop optical response with respect to PZT drive signal, $\hVLF/\Vinj$, is determined. The ratio of $\hVLF/\Vinj$ and Eq.~\ref{eq:mich_response} then yields the transfer function of the PZT drive, $\sop{M}\tu{inj}/\Vinj$.

Because the mechanical PZT resonances slowly drift in frequency over time (see \S\ref{sec:pzt_drive}), the PZT drives are recalibrated at approximately weekly intervals during data collection. Although larger drifts occur over longer periods, the PZT calibrations have been found to drift $<2\%$ from week to week. The time scale of the calibration drift is thus well-resolved by the weekly frequency of recalibration, bounding any residual biasing between recalibrations.

\subsection{Signal Readout Calibration $\,H^{\rm DAQ}_{\rm HF}(f)/H^{\rm DAQ}_{\rm LF}(f_0)$}
	\label{sec:signal_readout_transfer}

The direct, continuously-measured 1-kHz interferometer calibrations are transferred to higher frequencies by the signal-readout transfer functions, $\HdaqHF(f)/\HdaqLF(f_0)$. The large separation in scale between kHz and MHz frequencies requires a combination of different measurement techniques to measure these transfer functions over the entire range. Two measurements split at 900~kHz, a location determined by practical limitations of the measuring equipment. The following sections sequentially describe each measurement technique, starting at DC and progressing upward in frequency. 

\subsubsection{Measurement Below 900 kHz}
	\label{sec:lockin_technique}

Below 900~kHz, the signal-readout transfer functions are measured via an {\it ex situ} lock-in detection technique. Light from a 960nm LED is focused onto each modified New Focus 1811 photodetector (see \S\ref{sec:modified_detectors}), in turn. The LED power is modulated via a swept-sine signal, $\Vinj$, driven by a function generator. Averaged over many sweeps, the AC- and DC-coupled detector channels measure the signals
\begin{align}
\hVac &= \left(\HdaqHF(f) \; \Galign \; \Hled(f)\right) \; \Vinj \\
\hVdc &= \left(\HdaqLF(f) \; \Galign \; \Hled(f)\right) \; \Vinj
\end{align}
respectively. In the above equations, $\Galign$ is an alignment-dependent gain factor and $\Hled(f)$ is the unknown coupling of drive signal to optical power emitted by the LED. To eliminate the dependency on the unknown coupling function, the LED light is split by a beamsplitter and additionally directed onto a Thorlabs PDA 20CS photodetector, whose response, $\Href$, is known from external calibrations to be flat from DC to 1~MHz. Averaged over many sweeps, this second detector measures the signal
\begin{align}
&\hVref = \left( H^\mathrm{ref} \; {\Galign\tb{ref}} \; \Hled(f) \right) \; \Vinj
\end{align}
where $\Galign\tb{ref}$ is its alignment-dependent gain factor.

In terms of the measured sensor voltages, each signal-readout transfer function 
is
\begin{align}
\frac{\HdaqHF(f)}{\HdaqLF(f_0)}
 &= \left( \frac{\CSDjr[; f]{\hVac}{\Vinj}}{\CSDjr[; f]{\hVref}{\Vinj}} \right)
\left(  \frac{\CSDjr[; f_0]{\hVref}{\Vinj}}{\CSDjr[; f_0]{\hVdc}{\Vinj}} \right)
\end{align}
This measurement is independent of both optical alignment and LED output. Moreover, each of the response signals is ``locked-in'' to the coherent drive signal via CSDs computed by the high-speed DAQ system (see \S\ref{sec:correlator}), thus providing an unbiased measurement.

To accurately account for post-detection transmission losses, the LF and HF readout signals are both routed through their assigned {\it in situ} cabling from the photodetectors to the high-speed ADCs. The input load impedance of the ADCs, 50~$\Omega$ for the HF signals and 1~M$\Omega$ for the LF signals, is chosen to match the {\it in situ} measurement conditions. Although the LF signals are measured by the high-speed ADCs, rather than by the FPGAs, the difference is negligible at 1~kHz, because neither system applies signal conditioning other than a low-pass antialiasing filter. An additional 4-$7\%$ systematic error estimate is added in quadrature to the statistical uncertainty of each measurement. It accounts for a weak dependence of the responses of both the AC- and DC-coupled detector channels on the incident DC optical power. This effect is believed to arise from thermally-induced changes in resistances in the detector circuits at high power.

\subsubsection{Measurement above 900 kHz}

In principle, with broadband-sensing equipment, the lock-in technique described in the previous section could be used to calibrate the entire measurement band. However, the response of the secondary photodetector rolls off above 1~MHz, necessitating the use of a different gain reference for calibrating higher frequencies. A cut-off frequency of $f_1=900$~kHz is chosen as the point at which the response of the reference detector has rolled off by $1\%$ relative to its DC value. Above this frequency, the calibration can be decomposed into the transfer function product
\begin{align}
  \frac{\HdaqHF(f)}{\HdaqLF(f_0)} 
&=\left(\frac{\HdaqHF(f_1)}{\HdaqLF(f_0)}\right)
  \left(\frac{\HdaqHF(f)}{\HdaqHF(f_1)} \right)
\end{align}
where the previous lock-in measurement is used to bridge the DC-to-AC transfer from $f_0$ to $f_1$, followed by a second, AC-to-AC transfer from $f_1$ to higher frequencies. The gains and phases of the AC-to-AC transfer functions are measured separately, as described below.

\paragraph{Gain Measurement Using Optical Shot Noise}
	\label{sec:shot_noise_technique}

The gains of the AC-to-AC transfer functions are measured {\it ex situ} directly from optical shot noise. Light from an incandescent bulb, providing 0 to 150~mW of incident power, is focused onto each modified New Focus 1811 photodetector. The AC-coupled detector channel measures the signal
\begin{align}
\hVac &= \left( \HdaqHF(f) \sqrt{2 E_\lambda P\tb{inc}} \, \right) \W\tu{shot} + N\tu{dark}(f) \, \W\tu{dark}
\end{align}
where $E_\lambda$ is the average photon energy and $P\tb{inc}$ is the average power incident on the photodiode. The DAQ transfer function includes both the responsivity of the photodiode and the transimpedance gain of the detector circuit. The first noise mode, $\W\tu{shot}$, is the optical shot noise while the second mode, $\W\tu{dark}$, represents the dark noise of the detector. The power spectral density (PSD) of the signal,
\begin{align}
\label{eq:psd_shot_noise}
\PSD{\hVac; f} &= 2E_\lambda P\tb{inc} \abssq{\HdaqHF(f)}  + \abssq{N\tu{dark}(f)}
\end{align}
is biased by the presence of dark noise. Thus, the PSD is also measured at 0~mW of incident power,
\begin{align}
\PSD{\hVac; f}\tb{dark} &= \abssq{N\tu{dark}(f)}
\end{align}
and subtracted from Eq.~\ref{eq:psd_shot_noise}.

In terms of the measured sensor voltages, the AC-to-AC gain is
\begin{align}
\label{eq:ac_pd_gain}
\left|\frac{\HdaqHF(f)}{\HdaqHF(f_1)}\right|
  &= \sqrt{
    \frac
    {\PSDjr{\hVac; f} - \PSDjr{\hVac; f}\tb{dark}}
    {\PSDjr{\hVac; f_1} - \PSDjr{\hVac; f_1}\tb{dark}}
    }
\end{align}
where the PSDs are measured by the high-speed DAQ system via the {\it in situ} transmission cabling of each channel. The dark noise subtraction is checked across multiple tested incandescent powers and found to contribute an additional 1-$3\%$ systematic error (varying by frequency), which is added in quadrature to the statistical measurement error. These systematics again reflect a weak dependence of the response, and potentially the electronics noise (dark noise), of each modified detector on the incident DC optical power.

\paragraph{Phase Measurement Using Lock-In LED Drive}

While shot noise is useful as an inherently white source, it does not provide phase information. To measure phase at high frequency, the reference detector technique described in \S\ref{sec:lockin_technique} is modified to include a frequency-dependent response term, represented as $\Href(f)$. In terms of the measured sensor voltages, lock-in detection with the drive signal measures the phase response
\begin{align}
\label{eq:lockin_phase}
  \Arg{\frac{\HdaqHF(f)}{\HdaqHF(f_1)}} + 
  \Arg{\frac{\Href(f_1)}{\Href(f)}}
  \nonumber \\
&\hspace{-12em}= \Arg{ \left(\frac{\CSDjr[; f]{\Vac}{\Vinj}}{\CSDjr[; f]{\Vref}{\Vinj}}  \right)
  \left(\frac{\CSDjr[; f_1]{\Vref}{\Vinj}}{\CSDjr[; f_1]{\Vac}{\Vinj}} \right)}
\end{align}
where the reference detector responses at frequencies $f$ and $f_1$ do not cancel. As before, the CSDs are measured by
the high-speed DAQ system via the {\it in situ} transmission cabling of each channel. Eq.~\ref{eq:lockin_phase} is the
sum of the desired result and a phase offset due to the reference detector. Although the phase offset is not known, it
is common to the measurement of every signal-readout channel. It thus cancels inside the cross-spectrum between
channels, leaving only the phase difference between AC-to-AC transfer functions.

\subsection{Sampling Decoherence}
	\label{sec:timing_instability}

As discussed in \S\ref{sec:high_speed_adcs}, the high-speed ADC units operate under a distributed synchronization model. Each 100~MHz ADC sample clock is phase-locked to its own 1~MHz reference clock generated by a GPS timing card. The sample clocks freely run between every 100 samples, and the reference clocks themselves freely run over the 1~s duration between commonly-received GPS pulses. Relative drift of the sample clock rates between phase realignments causes decoherence in time-averaged measurements, resulting in a systematic degradation in sensitivity to correlated effects. This effect is characterized in the following sections.

\subsubsection{Measurement of Timing Stability}
	\label{sec:decoherence_measurement}

The degree of clock-rate drift between GPS-disciplined ADC units is measured by injecting stationary broadband noise from a common source into every channel. Because clock-rate drift occurs on a time scale $>1$~s, much longer than the 1-ms DFT duration, a contiguous time series of CSD measurements time-resolves the slow drift. To an individual CSD, the drift appears as a constant time offset, $\tau$, in the sampling of the two signals. The Fourier transform of the time series under a time shift $\tau$ is
\begin{align}
\dft{\sop{V}(t+\tau);f} &= e^{i 2 \pi \tau f} \dft{\sop{V}(t);f} 
\end{align}
The CSD of a signal and a time-shifted version of itself is thus
\begin{align}
\CSD{\sop{V}(t),\sop{V}(t + \tau);f} &= e^{-2\pi i \tau f} \; \PSD{\sop{V};f}
\end{align}
where the CSD phase depends purely on the sampling offset $\tau$. For each CSD in the time series, a least-squares fit of the parameter $\tau$ to the measured phase yields the instantaneous time offset between the sample clocks of the two channels. With multiple fits, the time dependence of $\tau$ itself shows the clock rate drift.

From this measurement, it is inferred that the internal control loop of each ADC allows its sample clock to freely drift
until hitting a rail at approximately $\pm 10$~ns, at which point a sudden frequency correction is made. This mechanism
suggests an asymptotically Gaussian distribution of clock offsets between two ADC chassis. An analogous measurement
establishes the two channels on the same ADC card/clock to be stable to $<1$ ns.
Fig.~\ref{fig:time_history_synchronicity} shows the time series of clock rate drift inferred via this method for
704~hours of recorded data.

\subsubsection{Clock Wander Decoherence Model}

The sampling time offsets shown in Fig.~\ref{fig:time_history_synchronicity} are approximately normally-distributed with a width of $\sigma_{\tau}\approx10$~ns. The aggregate statistical effect on a time-averaged CSD measurement can thus be modeled as if each CSD independently draws a random sampling time offset $\stau$ from a Gaussian distribution of variance $\sigma_\tau^2$. The decoherence of a time-averaged CSD measurement due to clocking noise is then
\begin{align}
\label{eq:decoherence_model}
Y(f) &= \big\langle e^{-i2\pi \sop{\tau} f} \big\rangle_{\tau}
= \mathrm{exp}\left(-2 \pi^2 \sigma_{\tau}^2 \, f^2\right)
\end{align}
In the presence of clocking phase noise, the measured noise power of intrinsically coherent signals is reduced by the decoherence function as
\begin{align}
\label{eq:decoherence_model_effect}
  \Big\langle \CSD{\sop{V}(t),\sop{V}(t + \sop{\tau})} \Big\rangle_{\tau}
  &= Y(f)  \; \PSD{\sop{V};f}
\end{align}
with $Y(f)\le 1$ at all frequencies.

Fig.~\ref{fig:decoherence} shows two measurements of the decohering of intrinsically coherent broadband noise (see \S\ref{sec:decoherence_measurement}), each computed from the time-averaged CSD after three hours of continuous integration. A fit of the Gaussian decoherence model (Eq.~\ref{eq:decoherence_model}) to the average of the two measurements yields a best-fit loss parameter of $\sigma_{\tau}=9.20$~ns. Because the Holometer signal bandwidth is less than 8~MHz, the inter-channel decorrelation due to clocking phase noise is $<10\%$ within this band. To account for the systematic loss of correlated sensitivity, the fitted decoherence function is applied as an overall gain correction to each of the calibrated cross-ADC CSD averages. 

\begin{figure}[!tp]
	\centering
    \includegraphics[width=1\linewidth]{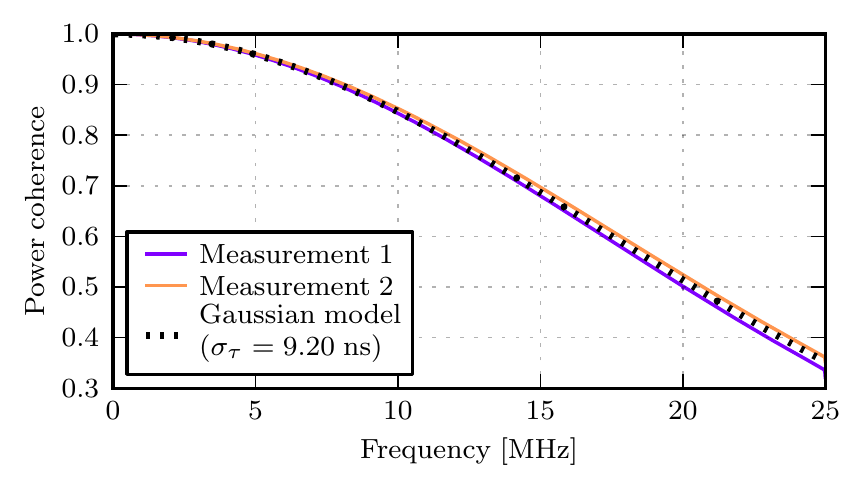}
    \caption[Decoherence of time-averaged CSD measurements due to clocking noise between the ADC units.]{
      Decoherence of time-averaged CSD measurements due to clocking noise between the ADC units. The purple and orange
      curves denote two three-hour measurements made via coherent broadband noise injection. The dashed black curve
      represents the fit of the Gaussian decoherence model (Eq.~\ref{eq:decoherence_model}) to the average of the two
      measurements. The loss of coherent signal is $<10\%$ below 10MHz.
    }
	\label{fig:decoherence}
\end{figure}

\section{Instrument Performance}
	\label{sec:performance}

The Holometer was operated over a six-month period, from July 2015 to February 2016, during which a total of 704~hours of data were acquired for scientific analysis. The following sections describe the sensitivity achieved by the instruments during this run and the systematic controls which ensure data quality. Using the notation conventions introduced in \S\ref{sec:Characterization}, subscripts specifying the readout channel will be dropped from all terms, but are implied. Instead, measurement estimators derived from data are subscripted with [square] bracketed labels. The discussion below applies to each channel.

\subsection{Contrast Defect}
  \label{sec:defect_performance}

When the interferometer is operated at a constant offset in AS transmissivity (see \ref{sec:darm_loop}), the contrast defect appearing in Eq.~\ref{eq:michelson_AS} diminishes the optical sensitivity to DARM modulations. This allows the calibration line dithers to be used as a reference to estimate the contrast defect. Substituting Eq.~\ref{eq:michelson_AS} into Eq.~\ref{eq:T_AS} gives an expression for the AS transmissivity, $T\tb{AS}$, which holds at all frequencies (by construction $T\tb{AS}$ has a flat, frequency-independent response). The derivative of this expression with respect to DARM displacement, $dT\tb{AS}/d\delta L$, thus corresponds to the magnitude of the measured calibration line dither at 1~kHz. Jointly solving $T\tb{AS}$ and $dT\tb{AS}/d\delta L$ for the contrast defect yields
\begin{align}
  \epsilon\tb{cd} = \frac{1}{2} - \sqrt{\frac{1}{4} - T\tb{AS} + T^2\tb{AS} + \left( \frac{\lambda}{4\pi} \frac{dT\tb{AS}}{d\delta L} \right)^2}
\end{align}
Since both $T\tb{AS}$ and $dT\tb{AS}/d\delta L$ are continuously-measured controls signals, the contrast defect can be
continually inferred for every time interval of recorded data.

Fig.~\ref{fig:defect_plot} shows the defect in the two interferometers around 100 hours into the first science observation run. It is representative of the typical defect. As shown, the sensitivity reliably changes over time after each lock to high power, suggesting that there is a thermal response affecting the contrast defect. In one of the two interferometers, the defect is substantially worse. In this instrument, the interference beam image suggests a mismatch of astigmatism, since it has an interference fringe profile similar to a percent sign (\%). This instrument is also the one requiring a dither signal in its alignment loop to break a degeneracy in the QPD signals (see \S\ref{sec:alignment_loop}). The smaller-defect instrument has an AS-port fringe profile resembling a bullseye, in which the outer ring becomes more prominent in camera profile images over the 30 seconds after the lock.

The ratio of fringe light to defect light affects the shot noise-limited sensitivity. The high-speed sensors were adjusted around hour 300 (shown in the next section) to improve this ratio, with an aperture cutting some of the defect light halo. This improved sensitivity in interferometer 2. The defects shown in Fig.~\ref{fig:defect_plot} are computed from the control AS-power pickoff sensors, which always view the whole AS beam profile.

\begin{figure}[!tp]
	\centering
    \includegraphics[width=1\linewidth]{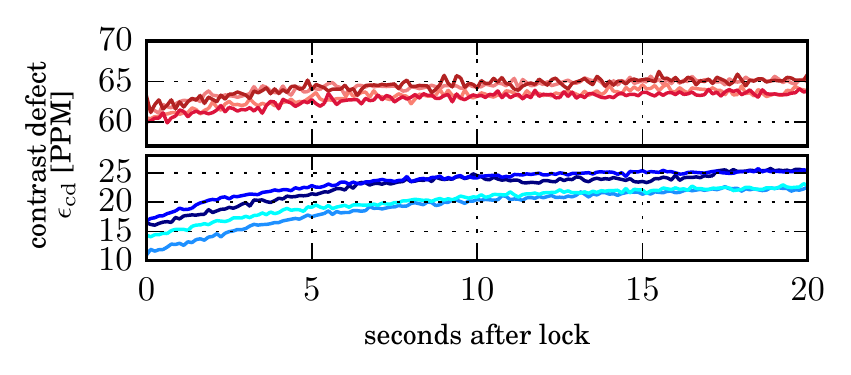}
    \caption{Contrast defect calculated from calibration-line data during the transition from initial lock acquisition
      to steady-state high-power operation. A series of lock acquisitions are shown by blue lines for interferometer 1
      and red lines for interferometer 2. These acquisitions occur near hour 100 in the 700-hour recorded data set. In both interferometers, the sensitivity reliably changes over time after each lock to high power, suggesting that there is a thermal response affecting the contrast defect.
    }
	\label{fig:defect_plot}
\end{figure}

\subsection{Actual Sensitivity}
	\label{sec:sensitivity}

Independently of the AC-coupled readout signal, which is used for scientific analysis, the DC-coupled readout signal of each photodetector provides a direct measure of the interferometer sensitivity. From time series measurements of the DC photovoltage, $\sop{V}\tu{DC}(t)$, and the 1 kHz calibration dither (see \S\ref{sec:low_freq_calib}), $\HcalLF(f_0,t)$, the estimator of the shot noise-limited sensitivity is
\begin{align}
\label{eq:dc_sls_estimator}
\left| N^{\rm shot}_{\rm [DC]}(f, t) \right|^2
	&= \frac{2e \left|\estZdc\right| \sop{V}\tu{DC}(t)}{\left|\HcalLF(f_0,t) \right|^2}
\end{align}
where $\left|\estZdc\right|=10$~V/A is the transimpedance gain of each DC-coupled channel. In the above equation, the numerator is the voltage noise power in the DC-coupled channel (units $\rm V^2/Hz$) and the denominator is the magnitude-square of the direct 1 kHz calibration (units $\rm V^2/m^2$).

Fig.~\ref{fig:time_history_sensitivity} shows the shot noise-limited sensitivity of each photodetector readout channel over the full integration time. The regular, step-like discontinuities in the channel sensitivities reflect periodic optical realignments of the inteferometers. These time histories demonstrate the interferometers to have maintained sensitivity to differential arm length displacements throughout the 700 hour integration time. Each interferometer achieves an average sensitivity of approximately $\rm 2\cdot10^{-18}\text{m}/\!\sqrt{\rm Hz}$.

\begin{figure}[!tp]
	\centering
    \includegraphics[width=1\linewidth]{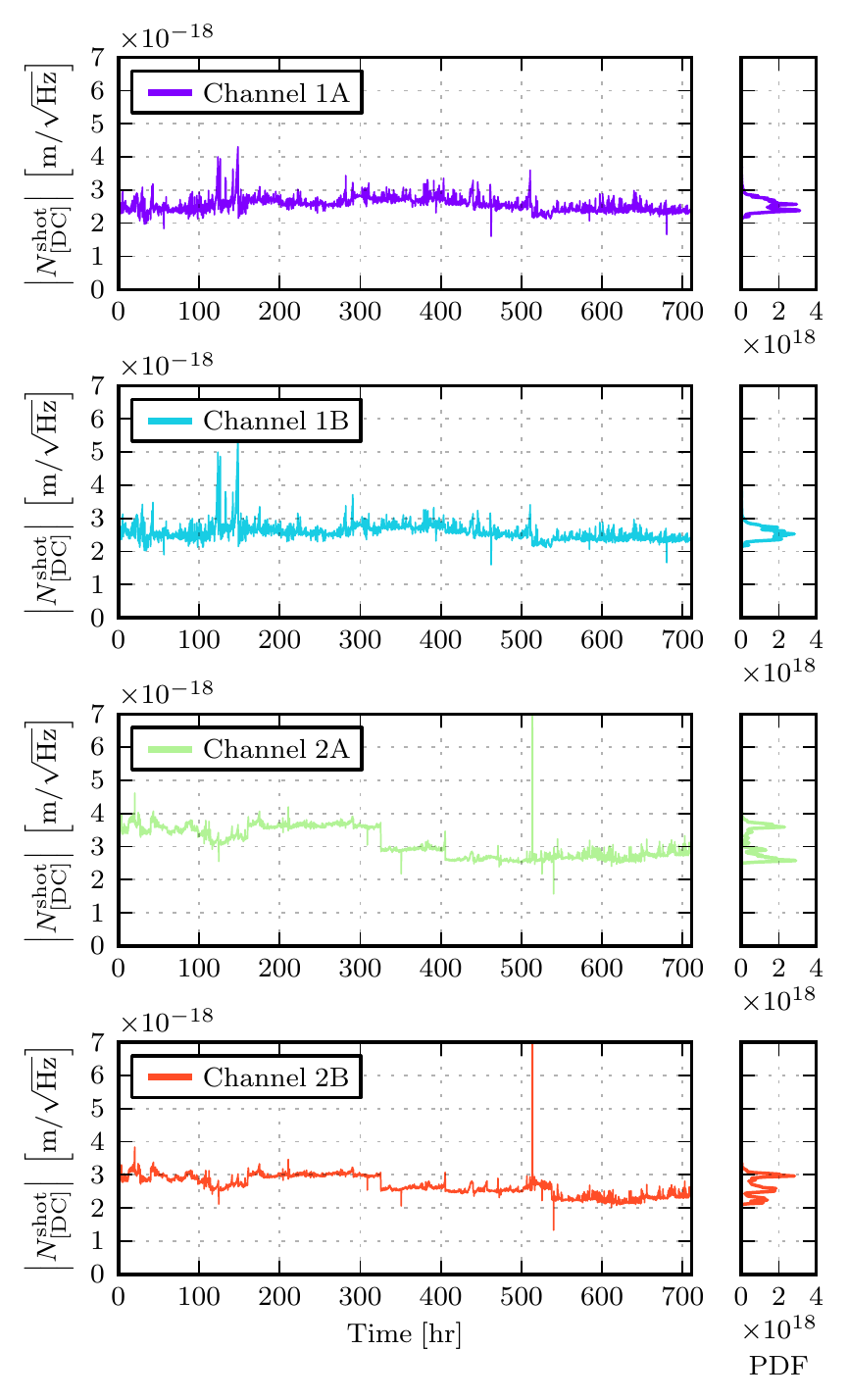}
    \caption[Time history of the shot noise-limited sensitivity of each interferometer channel.]{
      Time history of the shot noise-limited sensitivity of each interferometer readout channel. The regular, step-like discontinuities in the channel sensitivities reflect periodic realignments of the inteferometers during data collection. These time histories demonstrate the interferometers to have maintained sensitivity to differential arm length displacements throughout the 700-hour integration time. Each interferometer achieves an average sensitivity of approximately $\rm 2\cdot10^{-18}\text{m}/\!\sqrt{\rm Hz}$.
  }
	\label{fig:time_history_sensitivity}
\end{figure}

\subsection{Detector Stability}

The accuracy and stability of the calibrated photodetector responses can be tested {\it in situ} using the optical shot noise of the interferometers themselves. At a given shot noise-dominated frequency, chosen as $f_\text{SN}=2.3$~MHz, the diagonal elements of the calibrated 1-second CSD matrices are estimators of the instantaneous shot noise-limited sensitivity of each channel,
\begin{align}
\label{eq:ac_sls_estimator}
\left| N^{\rm shot}_{\rm [AC]}(f, t) \right|^2
	&= \PSD{\sop{M}^{\delta L};f_\text{SN},t}
\end{align}
If the photodetector response is not stable, the true transimpedance gains of its readout channels, $\left|\Zac\tb{[true]}(t)\right|$ and $\left|\Zdc\tb{[true]}(t)\right|$, will change in time relative to the values inferred from the {\it ex situ} signal-readout calibrations, $\left|\estZac\tb{[cal]}\right|$ and $\left|\estZdc\tb{[cal]}\right|$. Expanded in terms of its dependence on these detector gains, Eq.~\ref{eq:ac_sls_estimator} is
\begin{align}
\label{eq:ac_sls_estimator_decomposed}
\left| N^{\rm shot}_{\rm [AC]}(f, t) \right|^2
	&=
    \left(\frac{\left|\Zac\tb{[true]}(t)\right|^2
    \bigg/ \left|\estZac\tb{[cal]}\right|^2}{\left|\Zdc\tb{[true]}(t)\right|^2 \bigg/ \left|\estZdc\tb{[cal]}\right|^2}\right) \; \left| N^{\rm shot}_{\rm [true]}(f, t) \right|^2
\end{align}
where $\left| N^{\rm shot}_{\rm [true]}(f, t) \right|^2$ is the true shot noise-limited sensitivity. It can be seen that any deviation of the true detector gains from their {\it ex situ}-measured values biases the calibration-inferred experimental sensitivity.

As a cross-check against calibration bias, the calibration-inferred sensitivity (Eq.~\ref{eq:ac_sls_estimator}) can be compared to the direct estimator of the shot noise-limited sensitivity (Eq.~\ref{eq:dc_sls_estimator}), which depends only on the DC-channel readout signal. Expanded in terms of its dependence on the DC-channel detector gain, Eq.~\ref{eq:dc_sls_estimator} is
\begin{align}
\label{eq:dc_sls_estimator_decomposed}
\left| N^{\rm shot}_{\rm [DC]}(f, t) \right|^2
	&= \left(\left|\estZdc\tb{[cal]}\right| \bigg/ \left|\Zdc\tb{[true]}(t)\right|\right) \; \left| N^{\rm shot}_{\rm [true]}(f, t) \right|^2
\end{align}
Under this estimator, photodetector instability produces a {\it different} systematic bias than under Eq.~\ref{eq:ac_sls_estimator_decomposed}. The difference in estimator bias decouples photodetector stability from the intrinsically time-varying antisymmetric-port power. The ratio of Eqs.~\ref{eq:ac_sls_estimator_decomposed} and \ref{eq:dc_sls_estimator_decomposed} yields the test statistic
\begin{align}
\label{eq:pd_stability_statistic}
 R\tb{[SN]}(t) = \frac{\left| N^{\rm shot}_{\rm [AC]}(f, t) \right|^2}{\left| N^{\rm shot}_{\rm [DC]}(f, t) \right|^2}
	&= \frac{\left|\Zac\tb{[true]}(t)\right|^2 \bigg/ \left|\estZac\tb{[cal]}\right|^2}{\left|\Zdc\tb{[true]}(t)\right| \bigg/ \left|\estZdc\tb{[cal]}\right|}
\end{align}
representing the instantaneous fractional deviation of the true photodetector response ratio from that of the {\it ex situ} calibration measurement. If the photodetector response is stable, this statistic will equal a constant in time, and, if the {\it ex situ} calibration is accurate, then the value of this constant will equal unity. Eq.~\ref{eq:pd_stability_statistic} is sensitive to gain drift of either the AC- or DC-channel response individually and to common drift of both responses, through the quadratic versus linear dependence on the AC- and DC-channel gain biases, respectively.

Fig.~\ref{fig:time_history_pd_stability} shows the stability statistic of each photodetector over the full integration time. The flatness of the time histories are indicative of stable photodetector operation. The mean of each time history is consistent with unity to $<2\%$, indicating close agreement of the instantaneous detector responses with the {\it ex situ} calibrations. Moreover, the test statistic distributions, shown in the right panels, are all 5-$10\%$ in width, consistent with the total uncertainty inferred from the {\it ex situ} detector response measurements. This statistic thus provides evidence that the modified high-power photodetectors are both highly stable in their operation and well-characterized by the {\it ex situ} calibration measurements.

\begin{figure}[!tp]
	\centering
    \includegraphics[width=1\linewidth]{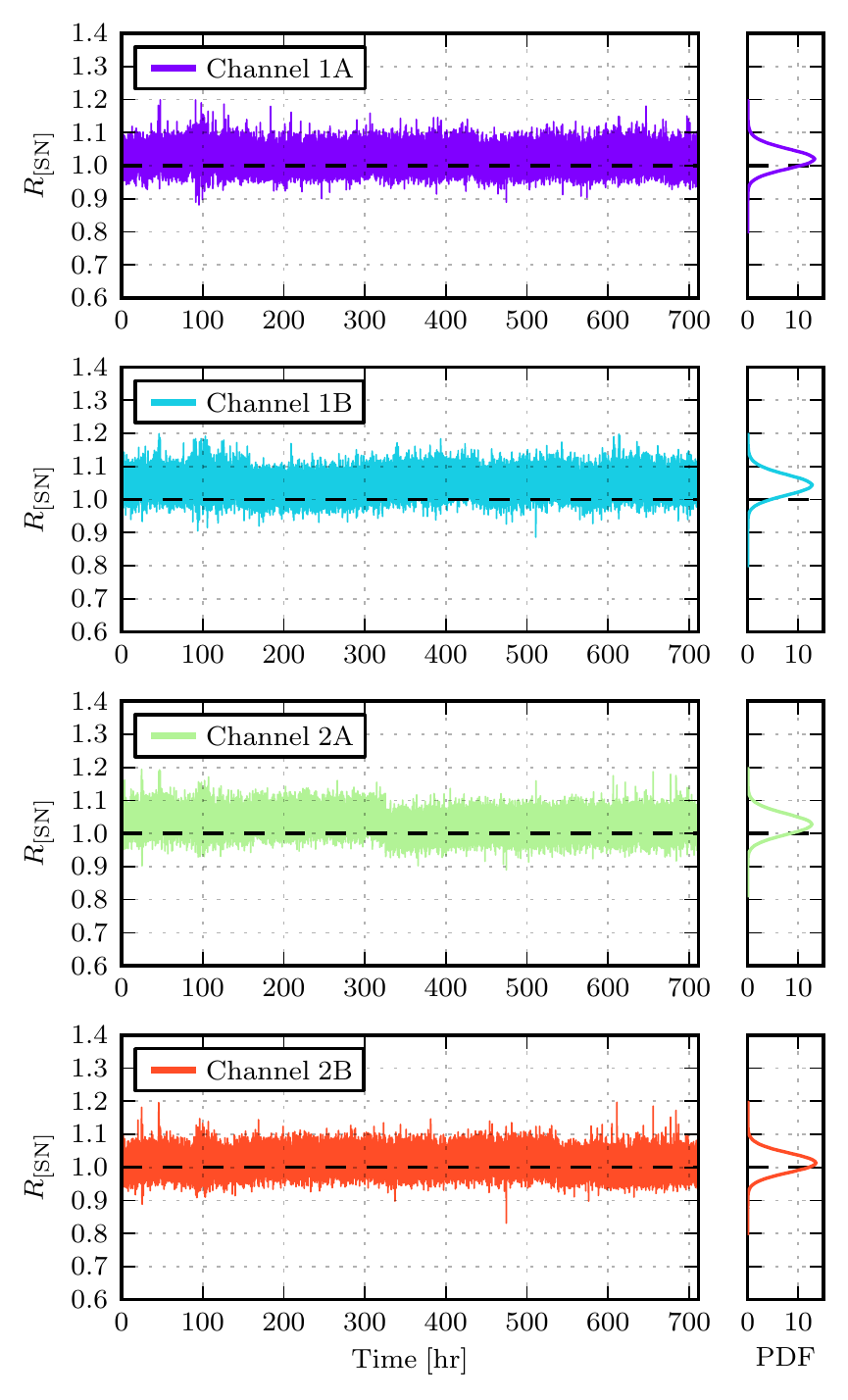}
    \caption[Time history of the photodetector transimpedance ratios, a proxy for response stability.]{
      Time history of the photodetector transimpedance ratios, a proxy for response stability. The flatness and near-unity values of these time histories are indicative of stable photodetector operation, with the instantaneous detector responses in close agreement with the {\it ex situ} calibrations. This provides compelling evidence that the modified high-power photodetectors are both highly stable in their operation and well-characterized by the {\it ex situ} calibrations.
    }
	\label{fig:time_history_pd_stability}
\end{figure}

\subsection{Sampling Synchronicity}
	\label{sec:synchronicity}

To ensure that correlated sensitivity is maintained throughout the 700-hour measurement duration, the sampling time
offset between the two ADC units receiving the AS-port detector readouts is continuously monitored. This is implemented via direct injection of a coherent narrowband optical signal onto the four high-power photodetectors at the antisymmetric port. A 960-nm LED installed in each output optics box is split by a beamsplitter and directed onto the two high-power detectors in each box. The 1-mW power of the two LEDs is coherently modulated via a function generator driving a sine wave at $13$~MHz. The instantaneous sampling time offset of each calibrated 1-s CSD matrix is inferred from the estimator
\begin{align}
\tau\tb{[LED]}(t)
	&= \frac{\text{Arg}\,\bigg[\CSD{\sop{M}\ifoL\michdl,\sop{M}\ifoT\michdl;f\tb{LED},t}\bigg]}{2 \pi f\tb{LED}}
\end{align}
evaluated at $f\tb{LED}=13$~MHz.

Fig.~\ref{fig:time_history_synchronicity} shows the time history of sampling drift between the two interferometer sum channels. As with the {\it ex situ} measurements (see \S\ref{sec:timing_instability}), the two ADC units remain tightly synchronized within a 10-ns envelope over the entire 700 hours. The fit of a normal distribution to the sample time offsets yields a best-fit width parameter of ${\sigma}_{\tau}=10.40$~ns, consistent with that of the {\it ex situ} decoherence measurements. In principle, the phase information could be used to re-phase each 1-s CSD matrix accumulation across the entire 700~hours, thus recovering the decoherence loss. However, in this analysis, the sensitivity loss is simply absorbed as a small ($<10\%$) gain correction applied to the final CSD measurement.

\begin{figure}[!tp]
	\centering
    \includegraphics[width=1\linewidth]{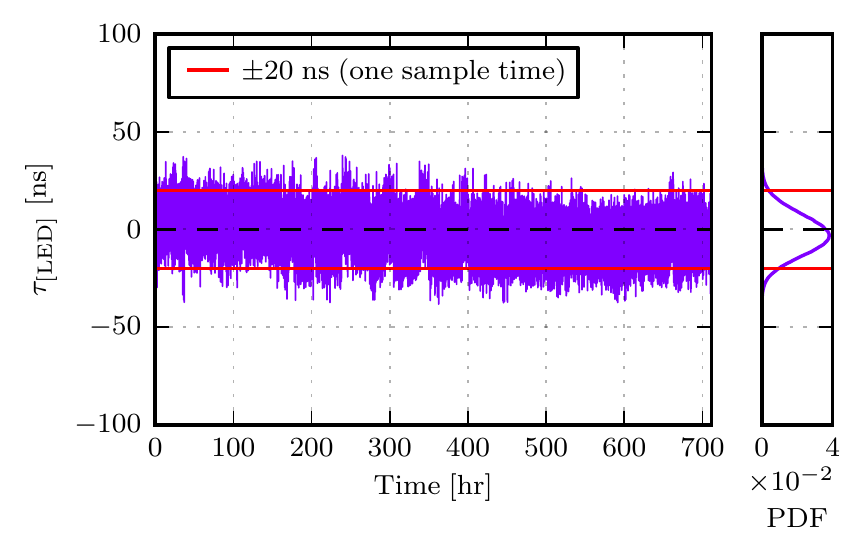}
    \caption[Time history of the sampling synchronicity between the two interferometer sum channels.]{
      Time history of the sampling synchronicity between the two interferometer sum channels as measured using an in
      situ 13MHz LED calibration. This time history demonstrates that the two ADC units remain tightly synchronized within a 10-ns envelope over the entire 700 hours. The fit of a normal distribution to the time offsets yields a best-fit width parameter of ${\sigma}_{\tau}=10.40$~ns and a mean of 2.17~ns, likely reflecting a small phase difference between the two separate sets of cabling connecting the LEDs to the function generator. This small offset corresponds to a phase difference of $<1^{\circ}$ at 1~MHz.
    }
	\label{fig:time_history_synchronicity}
\end{figure}

\subsection{Data Quality Control}
\label{sec:quality_control}

To ensure that the recorded data are sensitive to correlated position fluctuations and free from environmental contamination, a set of data quality standards are enforced. Data failing to meet these standards are excluded from analysis. Vetos related to loss of sensitivity are performed in the time domain, while those related to spurious environmental correlation are performed in the frequency domain. Each set of veto criteria is discussed below.

\subsubsection{Time-Resolved Vetos}

In the time domain, each 1-second accumulation of the CSD matrix is cross-validated against interferometer controls data from the same time interval. The controls data are used to confirm that the interferometers and their signal-readout electronics were operating in a stable, linear state at the time the CSD data were taken. The following set of veto conditions are enforced.

\paragraph{Out-of-Range Photodetector Power}

An optical exposure of 70-140~mW is required to be maintained on each photodetector. The nominal power incident on each detector is 100~mW. A power $<70$~mW is indicative of interferometer lock loss, while a power $>140$~mW exceeds the linear response range of the detector. The DC power, $P$, is inferred from the DC photovoltage, $V=RQP$, where $R=10$~V/A is the transimpedance gain of the DC channel and $Q=0.71$~A/W is the responsivity of the photodiode. 

\paragraph{Unstable Photodetector Response}

The response-stability statistic of every photodetector (Eq.~\ref{eq:pd_stability_statistic}) is required to remain within the range 0.8-1.2. The Gaussian width of this statistic is approximately 5\% for all channels, as shown in the right panels of Fig.~\ref{fig:time_history_pd_stability}, so excursions of $\pm 20\%$ are extreme statistical outliers. These outliers tend to coincide with interferometer lock loss, indicating that this veto is likely triggered by the detector signals ceasing to be shot noise-dominated, not by a true detector instability.

\paragraph{Loss of Sampling Synchronicity}

As described in \S\ref{sec:synchronicity}, the sampling synchronicity is monitored via a coherently-modulated LED directed onto the antisymmetric-port detectors. The coherence of the 13-MHz drive signal is required to be $\ge 0.9$ between all detectors. Loss of signal coherence also tends to conincide with lock loss/reacquisition, indicating that this veto is likely triggered by additional, transient noise diluting the drive coherence, not by a true synchronization loss.
\\\\%This newline is intentional for formatting
Vetos triggered by the above conditions are rare (${<}1\%$ of the data) and highly correlated with interferometer lock
loss. Because these vetos are all predicated on external operating conditions independent of the possible exotic noise, there is no sample biasing of the inferred noise constraints.

\subsubsection{Frequency-Resolved Vetos}

Frequencies determined to be contaminated by common environmental background are rejected from analysis. This is established through measurements of the cross-spectral densities between the antisymmetric-port detectors and the environmental noise monitors. Two examples of these measurements are the background limits on laser amplitude and phase noise, shown in Figs.~\ref{fig:ASLIM_IMON_lin} and \ref{fig:ASLIM_PDH_lin}, respectively. Frequencies at which the detectors exhibit significant ($>2\sigma$) correlation with the environmental monitors are identified as contaminated. As seen in the figures, the low frequencies are highly contaminated by correlated amplitude and phase noise from the lasers. On this basis, all frequency bins $<1$~MHz are vetoed from analysis. Because this veto is predicated on correlations with external signals insensitive to the quantum-geometrical noise, there is no sample biasing of the inferred noise constraints.

\section{Conclusion}
	\label{sec:conclusion}
	
This paper describes a new apparatus, the Fermilab Holometer, built to enable experimental studies of possible quantum-geometrical effects associated with new physics at the Planck scale. The device is currently being used to search for Planck-amplitude correlations of timelike trajectories in the classical macroscopic dimensions of space and time occupied by the apparatus. As a dedicated experiment, the Holometer is projected to provide Planck-precision tests of a wider class of quantum decoherence theories, significantly extending constraints from currently operating gravitational wave observatories.

\section*{Acknowledgements}

This work was supported by the Department of Energy at Fermilab under Contract No. DE-AC02-07CH11359 and the Early Career Research Program (FNAL FWP 11-03), and by grants from the John Templeton Foundation, the National Science Foundation (Grants No. PHY-1205254, No. DGE-0909667, No. DGE-0638477, and No. DGE-1144082), NASA (Grant No. NNX09AR38G), the Fermi Research Alliance, the Ford Foundation, the Kavli Institute for Cosmological Physics, University of Chicago/Fermilab Strategic Collaborative Initiatives, and the Universities Research Association Visiting Scholars Program. O.K. was supported by the Basic Science Research Program (Grant No. NRF-2016R1D1A1B03934333) of the National Research Foundation of Korea (NRF) funded by the Ministry of Education. The Holometer team gratefully acknowledges the extensive support and contributions of Bradford Boonstra, Benjamin Brubaker, Marcin Burdzy, Herman Cease, Tim Cunneen, Steve Dixon, Bill Dymond, Valera Frolov, Jose Gallegos, Emily Griffith, Hartmut Grote, Gaston Gutierrez, Evan Hall, Sten Hansen, Young-Kee Kim, Mark Kozlovsky, Dan Lambert, Scott McCormick, Erik Ramberg, Doug Rudd, Geoffrey Schmit, Alex Sippel, Jason Steffen, Sali Sylejmani, David Tanner, Jim Volk, William Wester, and James Williams for the design and construction of the apparatus.

\bibliographystyle{elsarticle-num}
\bibliography{holometer}

\begin{thebibliography}{10}
\expandafter\ifx\csname url\endcsname\relax
  \def\url#1{\texttt{#1}}\fi
\expandafter\ifx\csname urlprefix\endcsname\relax\def\urlprefix{URL }\fi
\expandafter\ifx\csname href\endcsname\relax
  \def\href#1#2{#2} \def\path#1{#1}\fi

\bibitem{Wheeler:1957}
J.~A. {Wheeler}, {On the nature of quantum geometrodynamics}, Annals of Physics
  2 (1957) 604--614.
\newblock \href {http://dx.doi.org/10.1016/0003-4916(57)90050-7}
  {\path{doi:10.1016/0003-4916(57)90050-7}}.

\bibitem{Hawking:1978}
S.~W. Hawking, \href{http://link.aps.org/doi/10.1103/PhysRevD.18.1747}{Quantum
  gravity and path integrals}, Phys. Rev. D 18 (1978) 1747--1753.
\newblock \href {http://dx.doi.org/10.1103/PhysRevD.18.1747}
  {\path{doi:10.1103/PhysRevD.18.1747}}.
\newline\urlprefix\url{http://link.aps.org/doi/10.1103/PhysRevD.18.1747}

\bibitem{Hawking:1980}
S.~Hawking, D.~Page, C.~Pope,
  \href{http://www.sciencedirect.com/science/article/pii/0550321380901510}{Quantum
  gravitational bubbles}, Nuclear Physics B 170~(2) (1980) 283 -- 306.
\newblock \href
  {http://dx.doi.org/http://dx.doi.org/10.1016/0550-3213(80)90151-0}
  {\path{doi:http://dx.doi.org/10.1016/0550-3213(80)90151-0}}.
\newline\urlprefix\url{http://www.sciencedirect.com/science/article/pii/0550321380901510}

\bibitem{Ashtekar:1992}
A.~Ashtekar, C.~Rovelli, L.~Smolin,
  \href{http://link.aps.org/doi/10.1103/PhysRevLett.69.237}{Weaving a classical
  metric with quantum threads}, Phys. Rev. Lett. 69 (1992) 237--240.
\newblock \href {http://dx.doi.org/10.1103/PhysRevLett.69.237}
  {\path{doi:10.1103/PhysRevLett.69.237}}.
\newline\urlprefix\url{http://link.aps.org/doi/10.1103/PhysRevLett.69.237}

\bibitem{Ellis:1992}
J.~{Ellis}, N.~E. {Mavromatos}, D.~V. {Nanopoulos}, {String theory modifies
  quantum mechanics}, Physics Letters B 293 (1992) 37--48.
\newblock \href {http://arxiv.org/abs/hep-th/9207103}
  {\path{arXiv:hep-th/9207103}}, \href
  {http://dx.doi.org/10.1016/0370-2693(92)91478-R}
  {\path{doi:10.1016/0370-2693(92)91478-R}}.

\bibitem{Hogan:2008a}
C.~J. {Hogan}, {Measurement of quantum fluctuations in geometry}, \prd 77~(10)
  (2008) 104031.
\newblock \href {http://arxiv.org/abs/0712.3419} {\path{arXiv:0712.3419}},
  \href {http://dx.doi.org/10.1103/PhysRevD.77.104031}
  {\path{doi:10.1103/PhysRevD.77.104031}}.

\bibitem{Hogan:2008b}
C.~J. {Hogan}, {Indeterminacy of holographic quantum geometry}, \prd 78~(8)
  (2008) 087501.
\newblock \href {http://arxiv.org/abs/0806.0665} {\path{arXiv:0806.0665}},
  \href {http://dx.doi.org/10.1103/PhysRevD.78.087501}
  {\path{doi:10.1103/PhysRevD.78.087501}}.

\bibitem{Hogan:2009}
C.~J. {Hogan}, M.~G. {Jackson}, {Holographic geometry and noise in matrix
  theory}, \prd 79~(12) (2009) 124009.
\newblock \href {http://arxiv.org/abs/0812.1285} {\path{arXiv:0812.1285}},
  \href {http://dx.doi.org/10.1103/PhysRevD.79.124009}
  {\path{doi:10.1103/PhysRevD.79.124009}}.

\bibitem{Hogan:2012}
C.~J. {Hogan}, {Interferometers as probes of Planckian quantum geometry}, \prd
  85~(6) (2012) 064007.
\newblock \href {http://arxiv.org/abs/1002.4880} {\path{arXiv:1002.4880}},
  \href {http://dx.doi.org/10.1103/PhysRevD.85.064007}
  {\path{doi:10.1103/PhysRevD.85.064007}}.

\bibitem{Hogan:2013}
C.~{Hogan}, {Quantum Geometry and Interferometry}, in: G.~{Auger},
  P.~{Bin{\'e}truy}, E.~{Plagnol} (Eds.), 9th LISA Symposium, Vol. 467 of
  Astronomical Society of the Pacific Conference Series, 2013, p.~17.
\newblock \href {http://arxiv.org/abs/1208.3703} {\path{arXiv:1208.3703}}.

\bibitem{Kwon:2014}
O.~{Kwon}, C.~J. {Hogan}, {Interferometric tests of Planckian quantum geometry
  models}, Classical and Quantum Gravity 33~(10) (2016) 105004.
\newblock \href {http://dx.doi.org/10.1088/0264-9381/33/10/105004}
  {\path{doi:10.1088/0264-9381/33/10/105004}}.

\bibitem{Hogan:2015a}
C.~J. {Hogan}, O.~{Kwon}, {Statistical Measures of Planck Scale Signal
  Correlations in Interferometers}, ArXiv e-prints\href
  {http://arxiv.org/abs/1506.06808} {\path{arXiv:1506.06808}}.

\bibitem{Hogan:2015b}
C.~{Hogan}, {Exotic Rotational Correlations in Quantum Geometry}, ArXiv
  e-prints\href {http://arxiv.org/abs/1509.07997} {\path{arXiv:1509.07997}}.

\bibitem{Hogan:2016}
C.~{Hogan}, O.~{Kwon}, J.~{Richardson}, {Statistical Model of Exotic Rotational
  Correlations in Emergent Space-Time}, ArXiv e-prints\href
  {http://arxiv.org/abs/1607.03048} {\path{arXiv:1607.03048}}.

\bibitem{Adhikari:2014}
R.~X. {Adhikari}, {Gravitational radiation detection with laser
  interferometry}, Reviews of Modern Physics 86 (2014) 121--151.
\newblock \href {http://arxiv.org/abs/1305.5188} {\path{arXiv:1305.5188}},
  \href {http://dx.doi.org/10.1103/RevModPhys.86.121}
  {\path{doi:10.1103/RevModPhys.86.121}}.

\bibitem{HoloPRL}
A.~S. Chou, R.~Gustafson, C.~Hogan, B.~Kamai, O.~Kwon, R.~Lanza, L.~McCuller,
  S.~S. Meyer, J.~Richardson, C.~Stoughton, R.~Tomlin, S.~Waldman, R.~Weiss,
  \href{http://link.aps.org/doi/10.1103/PhysRevLett.117.111102}{First
  measurements of high frequency cross-spectra from a pair of large michelson
  interferometers}, Phys. Rev. Lett. 117 (2016) 111102.
\newblock \href {http://dx.doi.org/10.1103/PhysRevLett.117.111102}
  {\path{doi:10.1103/PhysRevLett.117.111102}}.
\newline\urlprefix\url{http://link.aps.org/doi/10.1103/PhysRevLett.117.111102}

\bibitem{kogelnik_li}
H.~Kogelnik, T.~Li, Laser beams and resonators, Proceedings of the IEEE 54~(10)
  (1966) 1312--1329.
\newblock \href {http://dx.doi.org/10.1109/PROC.1966.5119}
  {\path{doi:10.1109/PROC.1966.5119}}.

\bibitem{Black2001}
E.~D. Black, {An introduction to Pound--Drever--Hall laser frequency
  stabilization} (2001).
\newblock \href {http://dx.doi.org/10.1119/1.1286663}
  {\path{doi:10.1119/1.1286663}}.

\bibitem{Loriette:2003}
V.~{Loriette}, C.~{Boccara}, {Absorption of Low-Loss Optical Materials Measured
  at 1064 nm by a Position-Modulated Collinear Photothermal Detection
  Technique}, Applied Optics 42 (2003) 649--656.
\newblock \href {http://dx.doi.org/10.1364/AO.42.000649}
  {\path{doi:10.1364/AO.42.000649}}.

\bibitem{Winkler:1991}
W.~{Winkler}, K.~{Danzmann}, A.~{R{\"u}diger}, R.~{Schilling}, {Heating by
  optical absorption and the performance of interferometric gravitational-wave
  detectors}, Physical Review A 44 (1991) 7022--7036.
\newblock \href {http://dx.doi.org/10.1103/PhysRevA.44.7022}
  {\path{doi:10.1103/PhysRevA.44.7022}}.

\bibitem{Freise:2004}
A.~{Freise}, G.~{Heinzel}, H.~{L{\"u}ck}, R.~{Schilling}, B.~{Willke},
  K.~{Danzmann}, {Frequency-domain interferometer simulation with higher-order
  spatial modes}, Classical and Quantum Gravity 21 (2004) S1067--S1074.
\newblock \href {http://arxiv.org/abs/gr-qc/0309012}
  {\path{arXiv:gr-qc/0309012}}, \href
  {http://dx.doi.org/10.1088/0264-9381/21/5/102}
  {\path{doi:10.1088/0264-9381/21/5/102}}.

\bibitem{flanagan:1994}
E.~E. {Flanagan}, K.~S. {Thorne}, {Noise Due to Backscatter Off Baffles, the
  Nearby Wall, and Objects at the Far End of the Beam Tube; and Recommended
  Actions}, Technical Report LIGO-T940063-00-R, LIGO (1994).

\bibitem{epics}
L.~R. Dalesio, J.~O. Hill, M.~Kraimer, S.~Lewis, D.~Murray, S.~Hunt, W.~Watson,
  M.~Clausen, J.~Dalesio, The experimental physics and industrial control
  system architecture: past, present, and future, Nuclear Instruments and
  Methods in Physics Research Section A: Accelerators, Spectrometers, Detectors
  and Associated Equipment 352~(1) (1994) 179--184.

\bibitem{medm}
K.~Evans~Jr, Medm reference manual, Argonne National Laboratory, http://www.
  aps. anl. gov/epics/EpicsDocumentation/ExtensionsManuals/MEDM/MEDM. html.

\bibitem{Welch1967}
P.~Welch, The use of fast fourier transform for the estimation of power
  spectra: A method based on time averaging over short, modified periodograms,
  IEEE Transactions on Audio and Electroacoustics 15~(2) (1967) 70--73.
\newblock \href {http://dx.doi.org/10.1109/TAU.1967.1161901}
  {\path{doi:10.1109/TAU.1967.1161901}}.

\bibitem{hwloc}
F.~Broquedis, J.~Clet-Ortega, S.~Moreaud, N.~Furmento, B.~Goglin, G.~Mercier,
  S.~Thibault, R.~Namyst, \href{https://hal.inria.fr/inria-00429889}{{hwloc: a
  Generic Framework for Managing Hardware Affinities in HPC Applications}}, in:
  IEEE (Ed.), {PDP 2010 - The 18th Euromicro International Conference on
  Parallel, Distributed and Network-Based Computing}, Pisa, Italy, 2010.
\newblock \href {http://dx.doi.org/10.1109/PDP.2010.67}
  {\path{doi:10.1109/PDP.2010.67}}.
\newline\urlprefix\url{https://hal.inria.fr/inria-00429889}

\bibitem{Harris1978}
F.~J. Harris, On the use of windows for harmonic analysis with the discrete
  fourier transform, Proceedings of the IEEE 66~(1) (1978) 51--83.
\newblock \href {http://dx.doi.org/10.1109/PROC.1978.10837}
  {\path{doi:10.1109/PROC.1978.10837}}.

\bibitem{schnupp}
L.~Schnupp, \href{http://relativity.livingreviews.org/refdb/record/23263}{Talk
  at european collaboration meeting on interferometric detection of
  gravitational waves}, Sorrento, Italy, 1988.
\newline\urlprefix\url{http://relativity.livingreviews.org/refdb/record/23263}

\end{thebibliography}

\end{document}
\endinput